\documentstyle [12pt,amsfonts] {article}
\input epsf

\newcommand{\beq}{\begin{equation}}
\newcommand{\eeq}{\end{equation}}
\newcommand{\beqs}{\begin{eqnarray}}
\newcommand{\eeqs}{\end{eqnarray}}

\catcode`@=11
\@addtoreset{equation}{section}
\@addtoreset{equation}{subsection}
\def\theequation{\ifnum\value{section}=0 \arabic{equation}\ignorespaces
\else \ifnum\value{section}=-1 A.\arabic{equation}\ignorespaces
\else \ifnum\value{subsection}=0 \thesection.\arabic{equation}\ignorespaces
\else \thesection.\arabic{subsection}.\arabic{equation}\ignorespaces
                           \fi
                      \fi
                 \fi}
\catcode`@=12

\begin{document}

\def\thefootnote{\fnsymbol{footnote}}

\baselineskip 6.0mm

\vspace{4mm}

\begin{center}
{\Large \bf General Structural Results for Potts Model Partition Functions on 
Lattice Strips} 

\vspace{8mm}

\setcounter{footnote}{0}
Shu-Chiuan Chang\footnote{email: shu-chiuan.chang@sunysb.edu} and
\setcounter{footnote}{6}
Robert Shrock\footnote{email: robert.shrock@sunysb.edu}

\vspace{6mm}

C. N. Yang Institute for Theoretical Physics  \\
State University of New York       \\
Stony Brook, N. Y. 11794-3840  \\

\vspace{10mm}

{\bf Abstract}
\end{center}

We present a set of general results on structural features of the $q$-state
Potts model partition function $Z(G,q,v)$ for arbitrary $q$ and temperature
Boltzmann variable $v$ for various lattice strips of arbitrarily great width
$L_y$ vertices and length $L_x$ vertices, including (i) cyclic and M\"obius
strips of the square and triangular lattice, and (ii) self-dual cyclic strips
of the square lattice.  We also present an exact solution for the chromatic
polynomial for the cyclic and M\"obius strips of the square lattice with width
$L_y=5$ (the greatest width for which an exact solution has been obtained so
far for these families).  In the $L_x \to \infty$ limit, we calculate the
ground-state degeneracy per site, $W(q)$ and determine the boundary ${\cal B}$
across which $W(q)$ is singular in the complex $q$ plane.

\vspace{16mm}

\pagestyle{empty}
\newpage

\pagestyle{plain}
\pagenumbering{arabic}
\renewcommand{\thefootnote}{\arabic{footnote}}
\setcounter{footnote}{0}

\section{Introduction} 

The $q$-state Potts model has served as a valuable model for the study of phase
transitions and critical phenomena \cite{potts,wurev}.  On a lattice, or, more
generally, on a (connected) graph $G$, at temperature $T$, this model is 
defined by the partition function
\beq
Z(G,q,v) = \sum_{ \{ \sigma_n \} } e^{-\beta {\cal H}}
\label{zfun}
\eeq
with the (zero-field) Hamiltonian
\beq
{\cal H} = -J \sum_{\langle i j \rangle} \delta_{\sigma_i \sigma_j}
\label{ham}
\eeq
where $\sigma_i=1,...,q$ are the spin variables on each vertex (site) 
$i \in G$;
$\beta = (k_BT)^{-1}$; and $\langle i j \rangle$ denotes pairs of adjacent
vertices.  The graph $G=G(V,E)$ is defined by its vertex set $V$ and its edge
set $E$; we denote the number of vertices of $G$ as $n=n(G)=|V|$ and the
number of edges of $G$ as $e(G)=|E|$.  We use the notation
\beq
K = \beta J \ , \quad a = e^K \ , \quad v = a-1
\label{kdef}
\eeq
so that the physical ranges are (i) $a \ge 1$, i.e., $v \ge 0$ corresponding to
$\infty \ge T \ge 0$ for the Potts ferromagnet, and (ii) $0 \le a \le 1$,
i.e., $-1 \le v \le 0$, corresponding to $0 \le T \le \infty$ for the Potts
antiferromagnet.  One defines the (reduced) free energy per site $f=-\beta F$,
where $F$ is the actual free energy, via
\beq
f(\{G\},q,v) = \lim_{n \to \infty} \ln [ Z(G,q,v)^{1/n}] 
\label{ef}
\eeq
where we use the symbol $\{G\}$ to denote $\lim_{n \to \infty}G$ for a given
family of graphs.  In the present context, this $n \to \infty$ limit
corresponds to the limit of infinite length for a strip graph of the square 
lattice of fixed width and some prescribed boundary conditions. 

Let $G^\prime=(V,E^\prime)$ be a spanning subgraph of $G$, i.e. a subgraph
having the same vertex set $V$ and an edge set $E^\prime \subseteq E$. Then
$Z(G,q,v)$ can be written as the sum \cite{fk}
\beq
Z(G,q,v) = \sum_{G^\prime \subseteq G} q^{k(G^\prime)}v^{e(G^\prime)}
\label{cluster}
\eeq
where $k(G^\prime)$ denotes the number of connected components of
$G^\prime$.  Since we only consider connected graphs $G$, we
have $k(G)=1$. The formula (\ref{cluster}) enables one to generalize $q$ from
${\mathbb Z}_+$ to ${\mathbb R}_+$ (keeping $v$ in its physical range).  The
formula (\ref{cluster}) shows that $Z(G,q,v)$ is a polynomial in $q$ and $v$.

The Potts model partition function is, up to a prefactor, equal to a quantity
of major current mathematical interest, the Tutte (also called Tutte/Whitney)
polynomial.  The Tutte polynomial of a graph $G$, $T(G,x,y)$, is given by
\cite{tutte1}-\cite{tutte3}
\beq
T(G,x,y)=\sum_{G^\prime \subseteq G} (x-1)^{k(G^\prime)-k(G)}
(y-1)^{c(G^\prime)}
\label{tuttepol}
\eeq
where $k(G^\prime)$, $e(G^\prime)$, and $n(G^\prime)=n(G)$ denote the number of
components, edges, and vertices of $G^\prime$, and $c(G^\prime) =
e(G^\prime)+k(G^\prime)-n(G^\prime)$ is the number of independent circuits in
$G^\prime$.  As stated in the text, $k(G)=1$ for the graphs of interest here.
Now let
\beq x=1+\frac{q}{v}
\label{xqv}
\eeq
and
\beq
y=a=v+1
\label{yqv}
\eeq
so that
\beq
q=(x-1)(y-1) \ .  
\label{qxy}
\eeq
Then
\beq
Z(G,q,v)=(x-1)^{k(G)}(y-1)^{n(G)}T(G,x,y) \ .
\label{zt}
\eeq

For a planar graph $G$ the Tutte polynomial satisfies the duality relation
\beq
T(G,x,y) = T(G^*,y,x)
\label{tuttedual}
\eeq
where $G^*$ is the (planar) dual to $G$.  Some reviews of Tutte polynomials
include \cite{tutte3}-\cite{boll}. 

One interesting special case of the Potts model is the zero-temperature limit
($v=-1$) of the Potts antiferromagnet, where the partition function is
identical to the chromatic polynomial $P(G,q)$ counting the number of ways of
coloring the vertices of a graph with $q$ colors subject to the condition that
no adjacent vertices have the same color \cite{birk,bbook,rrev,rtrev} 
\beq Z(G,q,-1)=P(G,q) \ . 
\label{zp}
\eeq
The minimum number of colors necessary for this coloring is the chromatic
number of $G$, denoted $\chi(G)$.  For sufficiently large $q$, on a given
lattice or graph $G$, the Potts antiferromagnet exhibits nonzero ground state
entropy (without frustration).  This is of interest as an exception to the
third law of thermodynamics \cite{al,cw}.  A physical example of residual
entropy at low temperatures is provided by ice \cite{lp}.  Nonzero ground state
entropy is equivalent to a ground state degeneracy per site (vertex), $W > 1$,
since $S_0 = k_B \ln W$.  We have
\beq
W(\{G\},q)= \lim_{n \to \infty}P(G,q)^{1/n} \ .
\label{w}
\eeq
A subtlety in the definition (\ref{w}) due to the noncommutativity 
\beq
\lim_{q \to q_s} \lim_{n \to \infty} P(G,q)^{1/n} \ne \lim_{n \to
\infty} \lim_{q \to q_s}P(G,q)^{1/n}
\label{wnoncom}
\eeq
at certain points $q_s$ (typically, $q=0,1,..,\chi(G)$) was discussed in
\cite{w}.

Using the formula (\ref{cluster}) for $Z(G,q,v)$, one can generalize $q$ from
${\mathbb Z}_+$ not just to ${\mathbb R}_+$ but to ${\mathbb C}$ and $a$ from
its physical ferromagnetic and antiferromagnetic ranges $1 \le a \le \infty$
and $0 \le a \le 1$ to $a \in {\mathbb C}$.  In particular, we shall be
interested here in the case of the $T=0$ antiferromagnet and the corresponding
zeros of $Z(G,q,-1)=P(G,q)$ (called chromatic zeros).  A subset of these zeros
can form a continuous accumulation set in the $n \to \infty$ limit, denoted
${\cal B}$.  This locus occurs where there is a non-analytic switching between
different $\lambda_{Z,G,j}$ of maximal magnitude and is thus determined as the
solution to the equation of degeneracy in magnitude of these maximal or
dominant $\lambda_{Z,G,j}$'s. As discussed earlier \cite{pm,bcc,a,tor4}, in the
infinite-length limit where the locus ${\cal B}$ is defined, for a given width
and transverse boundary condition (free or periodic) ${\cal B}$ depends on the
longitudinal boundary conditions but is independent of whether they involve
orientation reversal or not.  Thus, in the present context, for a given $L_y$,
${\cal B}$ is the same for the cyclic and M\"obius strips.  Following the
notation in \cite{w}, we denote the maximal region in the complex $q$ plane to
which one can analytically continue the function $W(\{G\},q)$ from physical
values where there is nonzero ground state entropy as $R_1$ .  The maximal
value of $q$ where ${\cal B}_q$ intersects the (positive) real axis is 
denoted $q_c(\{G\})$.  Thus, region $R_1$ includes the positive real axis for
$q > q_c(\{G\})$.

In the present work we shall give a number of general results on structural
features of the $q$-state Potts model partition function $Z(G,q,v)$ for
arbitrary $q$ and temperature Boltzmann variable $v$ for various lattice strips
of arbitrarily great width $L_y$ vertices and length $L_x$ vertices.  We shall
also discuss an exact solution for the chromatic polynomial for the cyclic and
M\"obius strips of the square lattice with width $L_y=5$ (the greatest width
for which an exact solution has been obtained so far for these families).  In
the $L_x \to \infty$ limit, we calculate the ground-state degeneracy per site,
$W(q)$ and determine the boundary ${\cal B}$ across which $W(q)$ is singular in
the complex $q$ plane.

There are several motivations for this work.  We have mentioned the basic
importance of nonzero ground state entropy in statistical mechanics.  From the
point of view of rigorous statistical mechanics, exact solutions are always
valuable for the insight that they give into the behavior of the given system
under study.  The results derived here give further insight into the general
structure of the Potts model partition function for lattice strips.  The study
of $W(\{G\},q)$ with $q$ generalized to complex values enables one to gain a
deeper understanding of the behavior of this function for $q \in {\mathbb
Z}_+$, in much the same way as the study of functions of a complex variable
imparts a deeper understanding of functions of a real variable in mathematics.
In particular, one sees how the value $q_c$ on the real axis corresponds to an
intersection of the locus ${\cal B}$ with this axis.  In addition to the papers
\cite{w,pm,bcc,a,tor4}, some relevant studies of Potts model partition
functions and/or chromatic polynomials for strip graphs of regular lattices
include \cite{lieb}-\cite{ts}.  Related mathematical papers, in addition to
those already mentioned, include \cite{read91}-\cite{brown}, and
Ref. \cite{ka3} contains further references.

\section{Structural Results}

\subsection{General}

In this section we briefly review some general structural results which will be
used for our new results below.  A general form for the Potts model partition
function for the strip graphs $G_m$ considered here, or more generally, for
recursively defined families of graphs comprised of $m$ repeated subunits
(e.g. the columns of squares of height $L_y$ vertices that are repeated $L_x=m$
times to form an $L_x \times L_y$ strip of a regular lattice with some
specified boundary conditions), is \cite{a}
\beq 
Z(G_m,q,v) = \sum_{j=1}^{N_{Z,G,\lambda}} c_{Z,G,j} 
(\lambda_{Z,G,j}(q,v))^m
\label{zgsum}
\eeq
where $N_{Z,G,\lambda}$, $c_{Z,G,j}$, and $\lambda_{Z,G,j}$ depend on the type
of recursive family $G$ (lattice structure and boundary conditions) but not on
its length $m$. For strips with periodic longitudinal boundary conditions
(e.g., cyclic and torus strips) eq. (\ref{zgsum}) follows from the property
that $Z(G_m,q,v)$ can be expressed as the trace of a certain transfer matrix,
\beq
Z(G_m,q,v)={\rm Tr}({\cal T}_Z^m) \ . 
\label{ztrace}
\eeq
Let
\beq
C_{Z,G} = \sum_{j=1}^{Z,G,\lambda} c_{Z,G,j} \ . 
\label{czsum}
\eeq
For the strips with periodic longitudinal boundary conditions, $C_{Z,G}$ is
thus the dimension of the matrix ${\cal T}_Z$.  In particular, for the cyclic
strip graphs of the square and triangular lattices considered here, the
repeated subgraph is the path graph consisting of $L_y$ vertices, and so,
denoting $C_{Z,G}=C_{Z,L_y}$ as before, one has \cite{cf}
\beq
C_{Z,L_y}=q^{L_y} \ . 
\label{czsumcyc}
\eeq
For strips with reversed-orientation periodic longitudinal boundary conditions
(M\"obius and Klein bottle boundary conditions), (\ref{ztrace}) is modified by
the insertion of the requisite orientation-reversal operator in the trace, with
the result that \cite{cf}
\beq
C_{Z,Mb,L_y}=q^{[\frac{L_y+1}{2}]}
\label{czsummb}
\eeq
where $[\nu]$ denotes the integral part of $\nu$. 

Equivalently, for the Tutte polynomial of a recursive graph 
\beqs
T(G_m,x,y) & = & \sum_{j=1}^{N_{T,G,\lambda}} c_{T,G,j}(q)
(\lambda_{T,G,j}(x,y))^m \cr\cr
& = & \frac{1}{x-1}\sum_{j=1}^{N_{T,G,\lambda}} \bar c_{T,G,j}(q)
(\lambda_{T,G,j}(x,y))^m
\label{tgsum}
\eeqs
where
\beq
\bar c_{T,G,j} = c_{Z,G,j} \ . 
\label{ctcz}
\eeq

For the special case of the $T=0$ antiferromagnet, the partition function, or
equivalently, the chromatic polynomial $P(G_m,q)$ has the corresponding form
\cite{bkw}
\beq
P(G_m,q) = \sum_{j=1}^{N_{P,G,\lambda}} c_{P,G,j} (\lambda_{P,G,j}(q))^m \ .
\label{pgsum}
\eeq
Let 
\beq
C_{P,G}=\sum_{j=1}^{N_{P,G,\lambda}} c_{P,G,j} \ . 
\label{cpsum}
\eeq
Here, the strip graph $G$ is indexed by its width, $L_y$. The analogue of 
(\ref{czsumcyc}) for the chromatic polynomial for cyclic strip
graphs of the square and triangular lattices is 
\beq
C_{P,L_y}=P(T_{L_y},q)=q(q-1)^{L_y-1}
\label{cpsumcyc}
\eeq
where $T_n$ denotes the tree graph with $n$ indices, and the analogue of 
(\ref{czsummb}) for the chromatic polynomial of the M\"obius strip of the
square lattice is \cite{cf}
\beq
C_{P,Mb,L_y} = \cases{ 0 & for even $L_y$ \cr
                       P(T_{(\frac{L_y+1}{2})},q) & for odd $L_y$ \cr } \
.
\label{cpsummb}
\eeq

\subsection{Cyclic and M\"obius Strips of the Square and Triangular Lattice} 

In \cite{cf} it was shown that for cyclic and M\"obius strips of the
square lattice of fixed width $L_y$ and arbitrary length $L_x$ (and also for
cyclic strips of the triangular lattice) the coefficients $c_j(q)$ in the Potts
model partition function (\ref{zgsum}), and hence, {\it a fortiori}, in the
chromatic polynomial (\ref{pgsum}), are polynomials in $q$ with the property
that there is a unique polynomial, denoted $c^{(d)}$, of degree $d$ in $q$. 
Further, this was shown to be a Chebyshev polynomial of the second kind:
\beq
c^{(d)} = U_{2d}(q^{1/2}/2) = \sum_{j=0}^d (-1)^j {2d-j \choose j} 
q^{d-j}
\label{cd}
\eeq
where $U_n(x)$ is the Chebyshev polynomial of the second kind.  A number of
properties of these coefficients were derived in \cite{cf}; for our present
work, we shall need the following special values: 
\beq
c^{(d)}=(-1)^d \quad {\rm for} \quad q=0
\label{cdq0}
\eeq
\beq
{\rm If} \quad q=1 \quad {\rm then} \quad
c^{(d)}= \cases{ 1 & if $d=0$ \ mod \ 3 \cr
                 0 & if $d=1$ \ mod \ 3 \cr
                -1 & if $d=2$ \ mod \ 3  }
\label{cdq1}
\eeq
\beq
{\rm If} \quad q=2 \quad {\rm then} \quad
c^{(d)}= \cases{ 1 & if $d=0,1$ \ mod \ 4 \cr
                 -1 & if $d=2,3$ \ mod \ 4 \cr } \ .
\label{cdq2}
\eeq
In contrast, for the M\"obius strips of the triangular lattice, the
coefficients in the Potts model partition function and chromatic polynomial do
not, in general, have the form $c^{(d)}$ \cite{wcy}.

Thus, the terms $\lambda_{Z,L_y,j}$ (equivalently,
$\lambda_{T,L_y,j}$) and $\lambda_{P,L_y,j}$ that occur in
(\ref{zgsum}), (\ref{tgsum}), and (\ref{pgsum}) can be classified into sets,
with the $\lambda_{Z,L_y,j}(q,v)$, $\lambda_{T,L_y,j}(x,y)$, and
$\lambda_{P,L_y,j}(q)$ in the $d$'th set being defined as those terms with
respective coefficients $c_{Z,L_y,j}=\bar c_{T,L_y,j}$ and $c_{P,L_y,j}$ being
equal to $c^{(d)}$.  In Ref. \cite{cf} the numbers of such terms, denoted
$n_Z(L_y,d)$ and $n_P(L_y,d)$ respectively, were calculated. Note that
\beq
n_Z(L_y,d)=n_T(L_y,d) \ .
\label{nznt}
\eeq
Labelling the eigenvalues with coefficient $c^{(d)}$ as $\lambda_{Z,L_y,d,j}$
with $1 \le j \le n_Z(L_y,d)$, the Potts model partition function can be
written in the form
\beq
Z(G[L_y \times m, cyc.],q,v) = \sum_{d=0}^{L_y} \bigg [
c^{(d)} \sum_{j=1}^{n_Z(L_y,d)} (\lambda_{Z,L_y,d,j})^m \bigg ]
\label{zgsumcyc}
\eeq
or equivalently, for the Tutte polynomial, 
\beq
T(G[L_y \times m, cyc.],x,y) = \frac{1}{x-1}\sum_{d=0}^{L_y} \bigg [
c^{(d)} \sum_{j=1}^{n_T(L_y,d)} (\lambda_{T,L_y,d,j})^m \bigg ]
\label{tgsumcyc}
\eeq
where 
\beq
\lambda_{Z,L_y,d,j}(q,v)=v^{L_y}\lambda_{T,L_y,d,j}(x,y)
\label{lamzlamt}
\eeq
with the relations (\ref{xqv}) and (\ref{yqv}). 

With analogous labelling, the chromatic polynomial has the form 
\beq
P(G[L_y \times m, cyc.],q) = \sum_{d=0}^{L_y} \bigg [
c^{(d)} \sum_{j=1}^{n_P(L_y,d)} (\lambda_{P,L_y,d,j})^m \bigg ] \ . 
\label{pgsumcyc}
\eeq

Combining (\ref{zgsumcyc}) with (\ref{czsumcyc}) and (\ref{pgsumcyc}) with
(\ref{cpsumcyc}), one has the following relations for the cyclic strip graphs
of the square and triangular lattice of width $L_y$ and arbitrary length
\cite{cf}
\beq
C_{Z,L_y} = \sum_{d=0}^{L_y} c^{(d)} n_Z(L_y,d) = q^{L_y}
\label{czsumcyccd}
\eeq
\beq
C_{P,L_y} = \sum_{d=0}^{L_y} c^{(d)} n_P(L_y,d) = q(q-1)^{L_y-1} \ . 
\label{cpsumcyccd}
\eeq

The total numbers, $N_{Z,L_y,\lambda}$ and
$N_{P,L_y,\lambda}$, of different terms $\lambda_{Z,L_y,j}$ and
$\lambda_{P,L_y,j}$ in eqs. (\ref{zgsum}) and (\ref{pgsum}) are given by
\cite{cf}
\beq
N_{Z,L_y,\lambda} = N_{T,L_y,\lambda} = \sum_{d=0}^{L_y} n_Z(L_y,d)
\label{nzsum}
\eeq
and
\beq
N_{P,L_y,\lambda} = \sum_{d=0}^{L_y} n_P(L_y,d) \ . 
\label{npsum}
\eeq

We shall recall here some results on the numbers $n_P(L_y,d)$; these are
nonzero for $0 \le d \le L_y$ and are given by 
\beq
n_P(L_y,L_y)=1
\label{npcly}
\eeq
\beq
n_P(1,0)=1
\label{np10}
\eeq
with all other numbers $n_P(L_y,d)$ being determined by the two recursion
relations \cite{cf} 
\beq
n_P(L_y+1,0)=n_P(L_y,1)
\label{nprecursion1}
\eeq
\beq
n_P(L_y+1,d) = n_P(L_y,d-1)+n_P(L_y,d)+n_P(L_y,d+1)
\quad {\rm for} \quad L_y \ge 1 \quad {\rm and} \quad 1 \le d \le L_y+1 \ .
\label{nprecursion2}
\eeq
The solution to these relations yielded, in particular, the results 
\beq
n_P(L_y,0)=M_{L_y-1}
\label{nplyd0}
\eeq
where $M_n$ is the Motzkin number, 
\beq
M_n =  \sum_{j=0}^n (-1)^j C_{n+1-j} {n \choose j}
\label{motzkin}
\eeq
where
\beq
C_n=\frac{1}{n+1}{2n \choose n}
\label{catalan}
\eeq
is the Catalan number.  Further, 
\beq
n_P(L_y,1)=M_{L_y} 
\label{nply1}
\eeq
and
\beq
n_P(L_y,L_y-1)=L_y \ . 
\label{nplym1}
\eeq
The total number of $\lambda_{Z,L_y,j}$'s in the Potts model partition function
$Z([L_y \times m,cyc.],q,v)$ or $Z([L_y \times m,Mb],q,v)$ is \cite{cf}
\beq
N_{Z,L_y,\lambda}={2L_y \choose L_y}
\label{nztot}
\eeq
and the total number of $\lambda_{P,L_y,j}$'s in the chromatic polynomial 
$P([L_y \times m,cyc.],q)$ or $P([L_y \times m,Mb],q)$ is \cite{cf}
\beq
N_{P,L_y,\lambda}=2(L_y-1)! \ \sum_{j=0}^{[\frac{L_y}{2}]} \frac{(L_y-j)}{
(j!)^2(L_y-2j)!} \ . 
\label{nptot}
\eeq

For arbitrary $L_y$, eq. (\ref{npcly}) shows that there is a unique 
$\lambda_{P,L_y,d}$ corresponding to the coefficient $c^{(d)}$ of highest 
degree, $d=L_y$, and this term is \cite{s4} 
\beq
\lambda_{P,L_y,d=L_y,j=1} = (-1)^{L_y} \equiv \lambda_{P,L_y,L_y} \ . 
\label{lamlyly}
\eeq
As indicated, since this eigenvalue is unique, it is not necessary to append a
third index, as with the other $\lambda$'s, and we avoid this for simplicity.
The result (\ref{lamlyly}) corresponds to the analogous property that in the
full Potts model partition function or equivalent Tutte polynomial,
$n_Z(L_y,L_y)=1$ \cite{cf} and the unique $\lambda_{Z,L_y,d=L_y}$ with
coefficient $c^{(L_y)}$ in $Z([L_y \times L_x,cyc.],q,v)$ or 
$Z([L_y \times L_x,Mb],q,v)$ is
\beq
\lambda_{Z,L_y,d=L_y}=v^{L_y} \ . 
\label{lamzlyly}
\eeq
Equivalently, the unique $\lambda_{T,L_y,d=L_y}$ with reduced coefficient $\bar
c_{T,G,j}=c^{(L_y)}$ in $T([L_y \times L_x,cyc.],x,y)$ or 
$T([L_y \times L_x,Mb],x,y)$ is 
\beq
\lambda_{T,L_y,d=L_y}=1 \ . 
\label{lamtlyly}
\eeq
(Here, one should not confuse the arguments of the Tutte polynomial $x,y$ with
the longitudinal and transverse directions $x,y$.)

For a strip of fixed width $L_y$ and arbitrary length $L_x=m$ with M\"obius 
boundary conditions, $(FBC_y,TPBC_x)$, the general form of the Potts model
partition function was determined in \cite{cf} to be  
\beq
Z(G[L_y \times m, Mb],q,v) = \sum_{d=0}^{d_{max,Mb}} 
\bigg \{ c^{(d)} \sum_{\eta=\pm 1} \eta \bigg [ 
\sum_{j=1}^{n_{Z,Mb}(L_y,d,\eta)} 
(\lambda_{Z,L_y,d,\eta,j})^m \bigg ] \bigg \} 
\label{zgsummb}
\eeq
where
\beq
 d_{max,Mb}=  \cases{ \frac{L_y}{2} & for even $L_y$ \cr
\frac{(L_y+1)}{2} & for odd $L_y$ \cr } \ .
\label{dmax}
\eeq
Correspondingly, 
\beq
T(G[L_y \times m, Mb],x,y) = \frac{1}{x-1}\sum_{d=0}^{d_{max,Mb}}
\bigg \{ c^{(d)} \sum_{\eta=\pm 1} \eta \bigg [ 
\sum_{j=1}^{n_{T,Mb}(L_y,d,\eta)}
(\lambda_{T,L_y,d,\eta,j})^m \bigg ] \bigg \}
\label{tgsummb}
\eeq
with $n_{T,Mb}(L_y,d,\eta)=n_{Z,Mb}(L_y,d,\eta)$.  For the special case of the
chromatic polynomial,
\beq
P(G[L_y \times m, Mb],q) = \sum_{d=0}^{d_{max,Mb}}
\bigg \{ c^{(d)} \sum_{\eta=\pm 1} \eta \bigg [ 
\sum_{j=1}^{n_{P,Mb}(L_y,d,\eta)}
(\lambda_{Z,L_y,d,\eta,j})^m \bigg ] \bigg \} \ . 
\label{pgsummb}
\eeq
Combining (\ref{zgsummb}) and (\ref{pgsummb}) with these formulas, one has the
relations for the M\"obius strip graphs of the square and triangular lattice of
width $L_y$ and arbitrary length \cite{cf} 
\beq
C_{Z,Mb,L_y} = \sum_{d=0}^{d_{max,Mb}} c^{(d)} \sum_{\eta=\pm 1} 
\eta n_{Z,Mb}(L_y,d,\eta) = q^{[\frac{L_y+1}{2}]} 
\label{czsummbcd}
\eeq
\beq
C_{P,Mb,L_y} = \sum_{d=0}^{d_{max,Mb}} c^{(d)} \sum_{\eta=\pm 1} \eta 
n_{P,Mb}(L_y,d,\eta) = \cases{ 0 & for even $L_y$ \cr
                       P(T_{(\frac{L_y+1}{2})},q) & for odd $L_y$ \cr } \
.
\label{cpsummbcd}
\eeq
In \cite{cf} the $n_{Z,Mb}(L_y,d,\eta)$ and $n_{P,Mb}(L_y,d,\eta)$ were
determined and a set of transformations were given that relate the Potts model
partition functions for the cyclic and M\"obius strips.

\subsection{Self-Dual Cyclic Strips of the Square Lattice} 

Consider a strip of the square lattice with (i) a fixed transverse width $L_y$,
(ii) an arbitrarily great length $L_x$, (iii) periodic longitudinal boundary
conditions, such that (iv) each vertex on one side of the strip, which we take
to be the upper side (with the strip oriented so that the longitudinal, $x$
direction is horizontal) are joined by edges to a single external vertex.  A
strip graph of this type will be denoted generically as $G_D$ (where the
subscript $D$ refers to the self-duality) and, when its size is indicated, as
$G_D(L_y \times L_x)$.  Although this family of graphs differs from the simple
recursive cyclic or M\"obius lattice strips, owing to the feature that all of
the vertices on the upper side are connected to a single external vertex, we
showed \cite{dg,sdg} that the Potts model partition function and its special
case, the chromatic polynomial, for the $G_D(L_y \times L_x)$ family have a
structure analogous to that for the cyclic and M\"obius strips of the square
lattice and cyclic strip of the triangular lattice, in the sense that there is
a unique coefficient which is a polynomial of degree $d$ in $q$, denoted
$\kappa^{(d)}$.  There are $n_{Z,G_D}(L_y,d)$ terms $\lambda_{Z,G_D,L_y,d,j}$
in $Z(G_D(L_y \times L_x),q,v)$ having coefficient $\kappa^{(d)}$.  The Potts
model partition function, Tutte polynomial, and chromatic polynomial were shown
to have the following forms \cite{dg,sdg}
\beq
Z(G_D[L_y \times L_x],q,v) = \sum_{d=1}^{L_y+1} \bigg [
\kappa^{(d)} \sum_{j=1}^{n_{Z,G_D}(L_y,d)} (\lambda_{Z,G_D,L_y,d,j})^m \bigg ]
\label{zgsumgd}
\eeq
or equivalently, for the Tutte polynomial, 
\beq
T(G_D[L_y \times m],x,y) = \sum_{d=1}^{L_y+1} \bigg [
\bar\kappa^{(d)} \sum_{j=1}^{n_{T,G_D}(L_y,d)} (\lambda_{T,G_D,L_y,d,j})^m 
\bigg ]
\label{tgsumgd}
\eeq
where
\beq
n_{T,G_D}(L_y,d)=n_{Z,G_D}(L_y,d)
\label{ntnz}
\eeq
and
\beqs
\kappa^{(d)} & = & q\bar\kappa^{(d)} \cr\cr
& = & \sqrt{q} \ U_{2d-1} \Big ( \frac{\sqrt{q}}{2} \Big ) \cr\cr
& = & \sum_{j=0}^{d-1} (-1)^j { 2d-1-j \choose j} q^{d-j} \cr\cr
& = & c^{(d)}+c^{(d-1)}
\label{kappad}
\eeqs
where $U_n(z)$ is the Chebyshev polynomial of the second kind. Thus,
$\kappa^{(1)}=q$, $\kappa^{(2)}=q(q-2)$, $\kappa^{(3)}=q(q-1)(q-3)$, and so
forth for higher values of $d$.  From (\ref{zt}) it follows that
\beq
\lambda_{Z,G_D,L_y,d,j}(q,v)=v^{L_y}\lambda_{T,G_D,L_y,d,j}(x,y) \ , \quad 1
\le d \le L_y + 1
\label{lamzlamtgd}
\eeq
with the relations (\ref{xqv}) and (\ref{yqv}).  Clearly, for the total
numbers, $N_{Z,G_D,L_y,\lambda}=N_{T,G_D,L_y,\lambda}$. 

For the chromatic polynomial, 
\beq
P(G_D[L_y \times L_x],q) = \sum_{d=1}^{L_y+1} \bigg [
\kappa^{(d)} \sum_{j=1}^{n_{P,G_D}(L_y,d)} (\lambda_{P,G_D,L_y,d,j})^m \bigg ]
\ . 
\label{pgsumgd}
\eeq

The analogues to (\ref{czsumcyccd}) and (\ref{cpsumcyccd}) for this self-dual 
strip of the square lattice are
\beq
\sum_{d=1}^{L_y+1} \kappa^{(d)} n_{Z,G_D}(L_y,d) = q^{L_y+1}
\label{czsumgd}
\eeq
\beq
\sum_{d=1}^{L_y+1} \kappa^{(d)} n_{P,G_D}(L_y,d) = P(T_{L_y+1},q) = 
q(q-1)^{L_y} \ . 
\label{cpsumgd}
\eeq

A number of properties of the $\kappa^{(d)}$ coefficients were derived in
\cite{dg}; for our present work, we use the following special values
\beq
\kappa^{(d)}=0  \quad {\rm for} \quad q=0
\label{kappadq0}
\eeq
\beq
{\rm If} \quad q=1 \quad {\rm then} \quad
\kappa^{(d)}= \cases{ 0 & if $d=0$ \ mod \ 3 \cr
                      1 & if $d=1$ \ mod \ 3 \cr
                     -1 & if $d=2$ \ mod \ 3  }
\label{kappadq1}
\eeq
\beq
{\rm If} \quad q=2 \quad {\rm then} \quad
\kappa^{(d)}= \cases{  0 & if $d$ \ is \ even \cr
                  2(-1)^k & if $d$ \ is \ odd \ and $d=2k+1$ \cr} \ .
\label{kappadq2}
\eeq

The numbers $n_{Z,G_D}(L_y,d)$ and $n_{P,G_D}(L_y,d)$ were determined in
\cite{dg} and related to $n_Z(L_y,d)$ and $n_P(L_y,d)$ for cyclic strips of the
square and triangular lattices.  For the total number of terms in the Potts
model partition function,
\beq
N_{Z,G_D,L_y,\lambda} = { 2L_y+1 \choose L_y+1} \ . 
\label{nztotgd}
\eeq
The analogous total number of terms in the chromatic polynomial is given by
\beq
\frac{1}{2}\biggl [ \biggl ( \frac{1+x}{1-3x} \biggr )^{1/2} - 1 \biggr ] - x =
\sum_{L_y=1}^\infty N_{P,G_D,L_y,\lambda}x^{L_y+1} 
\label{genfungd}
\eeq
and satisfies
\beq
N_{P,G_D,L_y,\lambda} = \frac{1}{2}N_{P,L_y+1,\lambda} 
\label{nptotrelgd}
\eeq
where the latter refers to cyclic or M\"obius strips of the square or
triangular lattice. 

\section{Structural Sum Rules} 

\subsection{Relations Between Sums of $n_Z(L_y,d)$}

In this section we derive new structural relations for the Potts model
partition function and chromatic polynomial.  These involve sums of terms and
hence, following a traditional nomenclature in physics, we shall refer to them
as sum rules.

\medskip

{\bf Proposition 1} \quad For the strip of the square or triangular lattice of
width $L_y$ and arbitrary length $L_x$ with $(FBC_y,PBC_x)$ (i.e., cyclic)
boundary conditions,
\beq
\sum_{\stackrel{0 \le d \le L_y}{d \ {\rm even}}} n_Z(L_y,d) = 
\sum_{\stackrel{0 \le d \le L_y}{d \ {\rm odd }}} n_Z(L_y,d) 
= \frac{1}{2}{2L_y \choose L_y} \ . 
\label{nzq0}
\eeq

\medskip

{\bf Proof} \quad Evaluating (\ref{czsumcyccd}) for $q=0$ and using
(\ref{cdq0}) yields the first equality.  From this equality and the fact that
the sum of the number of eigenvalues $\lambda_{Z,L_y,d,j}$ with coefficients of
even and odd $d$ is equal to (\ref{nztot}), the last equality in (\ref{nzq0})
follows. $\Box$

\medskip

Clearly, eq. (\ref{nzq0}) also applies with $n_Z(L_y,d)$ replaced by
$n_T(L_y,d)$.  Note that the right-hand side of the last equality in
(\ref{nzq0}) is equal to the number of directed animals of length $n=L_y$
on the triangular lattice \cite{cf}.  The first few values of this quantity for
$1 \le L_y \le 5$ are 1, 3, 10, 35, and 126, respectively.

\medskip

It is of interest to observe a related structural feature for cyclic strips of
the square or triangular lattice.  We start by noting that for any graph $G$,
$Z(G,q,v)=0$ if $q=0$, as is clear from (\ref{cluster}).  We apply this to the
case where $G$ is the cyclic strip of the square or triangular lattice, denoted
for short simply as $[L_y \times L_x,cyc.]$ with $L_x=m$, and use (\ref{cdq0})
to obtain
\beqs
& & Z([L_y \times m,cyc.],q=0,v) = 0 \cr\cr
& = & \sum_{\stackrel{0 \le d \le L_y}{d \ {\rm even}}}
\sum_{j=1}^{n_Z(L_y,d)} (\lambda_{Z,L_y,d,j}(0,v))^m 
- \sum_{\stackrel{0 \le d \le L_y}{d \ {\rm odd}}}
\sum_{j=1}^{n_Z(L_y,d)} (\lambda_{Z,L_y,d,j}(0,v))^m \ .  
\label{zq0}
\eeqs
We observe from our explicit calculations that the terms
$\lambda_{Z,L_y,d,j}(q,v)$ do not, in general, vanish at $q=0$. Since the
cancellation between the various $m$'th powers of the eigenvalues must occur
for arbitrary $m$, they must, in fact, cancel in pairs.  Note that this
provides a second proof of the result (\ref{nzq0}), since the cancellation
requires that the number of eigenvalues $\lambda_{Z,L_y,d,j}$ with even $d$
must be equal to the number of eigenvalues $\lambda_{Z,L_y,d,j}$ with odd $d$.

\medskip

Next, we have 

\medskip

{\bf Prop. 2} \quad For the cyclic strip of the square or triangular lattice
of width $L_y$ and arbitrary length $L_x$
\beq
\sum_{\stackrel{0 \le d \le L_y}{d=0 \ {\rm mod} \ 3}} n_Z(L_y,d) - 
\sum_{\stackrel{0 \le d \le L_y}{d=2 \ {\rm mod} \ 3}} n_Z(L_y,d) = 1 \ . 
\label{nzq1}
\eeq

\medskip

{\bf Proof} \quad To prove this, we evaluate (\ref{czsumcyccd}) for $q=1$ and
use (\ref{cdq1}).  $\Box$

\medskip

{\bf Prop. 3} \quad For the cyclic strip of the square or triangular lattice
of width $L_y$ and arbitrary length $L_x$
\beq
\sum_{\stackrel{0 \le d \le L_y}{d=0,1 \ {\rm mod} \ 4}} n_Z(L_y,d) -
\sum_{\stackrel{0 \le d \le L_y}{d=2,3 \ {\rm mod} \ 4}} n_Z(L_y,d) = 2^{L_y}
 \ . 
\label{nzq2}
\eeq

\medskip

{\bf Proof} \quad To prove this, we evaluate (\ref{czsumcyccd}) for $q=2$ and 
use (\ref{cdq2}).  $\Box$

\medskip

{\bf Prop. 4} \quad For the cyclic strip of the square or triangular lattice
of width $L_y$ and arbitrary length $L_x$
\beqs
\sum_{\stackrel{0 \le d \le L_y}{d=0,1 \ {\rm mod} \ 4}} n_Z(L_y,d) & = & 
\frac{1}{2}{2L_y \choose L_y} + 2^{L_y-1} \cr\cr
& & \cr\cr
\sum_{\stackrel{0 \le d \le L_y}{d=2,3 \ {\rm mod} \ 4}} n_Z(L_y,d) & = &
\frac{1}{2}{2L_y \choose L_y} - 2^{L_y-1} \ . 
\label{nzsum01mod423mod4}
\eeqs

\medskip

{\bf Proof} \quad  This result follows from (\ref{nzq2}) and the fact that the
sum $\sum_{0 \le d \le L_y} n_Z(L_y,d) = N_{Z,L_y,\lambda}$, for which one has
the expression given in eq. (\ref{nztot}).  $\Box$

\subsection{Relations Between Sums of $n_P(L_y,d)$}

\medskip

{\bf Prop. 5} \quad For the cyclic strip of the square or triangular 
lattice of width $L_y$ and arbitrary length $L_x$
\beqs
\sum_{\stackrel{0 \le d \le L_y}{d \ {\rm even}}} n_P(L_y,d) & = & 
\sum_{\stackrel{0 \le d \le L_y}{d \ {\rm odd }}} n_P(L_y,d) \cr\cr
& = & (L_y-1)! \ \sum_{j=0}^{[\frac{L_y}{2}]} \frac{(L_y-j)}{(j!)^2(L_y-2j)!} 
\ . 
\label{npq0}
\eeqs
\bigskip

{\bf Proof} \quad Evaluating (\ref{cpsumcyccd}) for $q=0$ and using
(\ref{cdq0}) yields the first equality.  From this equality and the fact that
the sum of the number of eigenvalues $\lambda_{P,L_y,d,j}$ with coefficients of
even and odd $d$ is equal to (\ref{nptot}), the last equality in (\ref{npq0})
follows.  $\Box$

\medskip

Note that the right-hand side of (\ref{npq0}) is equal to the number of
directed animals of length $L_y$ on the square lattice \cite{cf}.  The first
few values of this quantity for $1 \le L_y \le 5$ are 1, 2, 5, 13, and 35, 
respectively.

Again, one can remark on a related structural feature for cyclic strips of the
square or triangular lattice.  Clearly, $P(G,q)=0$ if $q=0$.  We apply this to
the case where $G$ is the cyclic strip of the square or triangular lattice,
denoted for short simply as $[L_y \times L_x,cyc.]$ with $L_x=m$, and use
(\ref{cdq0}) to obtain
\beqs
& & P([L_y \times m,cyc.],q=0) = 0 \cr\cr
& = & \sum_{\stackrel{0 \le d \le L_y}{d \ {\rm even}}}
\sum_{j=1}^{n_P(L_y,d)} (\lambda_{P,L_y,d,j}(q=0))^m 
- \sum_{\stackrel{0 \le d \le L_y}{d \ {\rm odd}}} 
\sum_{j=1}^{n_P(L_y,d)} (\lambda_{P,L_y,d,j}(q=0))^m \ . 
\label{pq0}
\eeqs
We observe from our calculations that the $\lambda_{P,L_y,d}$'s are not, in
general, zero for $q=0$. Since the cancellation between the various
$m$'th powers of the eigenvalues must occur for arbitrary $m$, they must, in
fact, cancel in pairs.  Hence, the number of eigenvalues $\lambda_{P,L_y,d,j}$
with even $d$ must be equal to the number of eigenvalues $\lambda_{P,L_y,d,j}$
with odd $d$. This yields another way of seeing the result (\ref{npq0}). 

\medskip

Next, 

\medskip

{\bf Prop. 6} \quad For the cyclic strip of the square or triangular
lattice of width $L_y \ge 2$ and arbitrary length $L_x$
\beqs
& & \sum_{\stackrel{0 \le d \le L_y}{d=0 \ {\rm mod} \ 3}} n_P(L_y,d) = 
\sum_{\stackrel{0 \le d \le L_y}{d=2 \ {\rm mod} \ 3}}     n_P(L_y,d) \cr\cr 
& = & \frac{1}{2}\biggl [ N_{P,L_y,d}-N_{P,L_y-1,d} \biggr ] \ . 
\label{npq1}
\eeqs
where the expression for $N_{P,L_y,d}$ was given in eq. (\ref{nptot}). 

\medskip

{\bf Proof} \quad To prove the first equality, we evaluate (\ref{cpsumcyccd})
for $q=1$ and use (\ref{cdq1}).  For the second equality, we observe that from
the results of \cite{cf},
\beq
\sum_{\stackrel{0 \le d \le L_y}{d = 1 \ {\rm mod} \ 3}}n_P(L_y,d) =
N_{P,L_y-1,\lambda}
\label{npsumd1mod3}
\eeq
where the expression for $N_{P,L_y,\lambda}$ was given in (\ref{nptot}). 
Together with the first equality in (\ref{npq1}) and the formula from \cite{cf}
for the total sum, (\ref{nptot}), this yields the second equality. $\Box$ 

\medskip

For $2 \le L_y \le 6$ the values of the right-hand side of the second equality
in (\ref{npq1}) are 1, \ 3, \ 8, \ 22, \ 61.  The case $L_y=1$ is special since
for this case $C_{P,L_y=1}=1$ rather than 0 at $q=1$, so that the analogue to
eq. (\ref{npq1}) reads simply $n_P(1,0)=1$.

\medskip

{\bf Prop. 7} \quad For the cyclic strip of the square or triangular lattice
of width $L_y$ and arbitrary length $L_x$
\beq
\sum_{\stackrel{0 \le d \le L_y}{d=0,1 \ {\rm mod} \ 4}} n_P(L_y,d) -
\sum_{\stackrel{0 \le d \le L_y}{d=2,3 \ {\rm mod} \ 4}} n_P(L_y,d) = 2 \ .
\label{npq2}
\eeq

\medskip

{\bf Proof} \quad To prove this, we evaluate (\ref{cpsumcyccd}) for $q=2$ and
use (\ref{cdq2}).  $\Box$

\medskip

{\bf Prop. 8} \quad For the cyclic strip of the square or triangular lattice
of width $L_y$ and arbitrary length $L_x$
\beqs
\sum_{\stackrel{0 \le d \le L_y}{d=0,1 \ {\rm mod} \ 4}} n_P(L_y,d) & = &
\frac{1}{2}N_{P,L_y,\lambda} + 1 \cr\cr
& & \cr\cr
\sum_{\stackrel{0 \le d \le L_y}{d=2,3 \ {\rm mod} \ 4}} n_P(L_y,d) & = &
\frac{1}{2}N_{P,L_y,\lambda} - 1 \ . 
\label{npsum01mod423mod4}
\eeqs

\medskip

{\bf Proof} \quad  This result follows from (\ref{npq2}) and the fact that the
sum $\sum_{0 \le d \le L_y} n_P(L_y,d) = N_{P,L_y,\lambda}$, for which one has
the expression given in eq. (\ref{nptot}).  $\Box$

\medskip

For an arbitrary connected graph with at least two vertices, $P(G,q)=0$ for
$q=1$.  It follows, in particular, that 
\beqs
& & P([L_y \times m,cyc.],q=1) = 0 \cr\cr
& = & \sum_{\stackrel{0 \le d \le L_y}{d=0 \ {\rm mod} \ 3}}
\sum_{j=1}^{n_P(L_y,d)} (\lambda_{P,L_y,d,j}(q=1))^m 
- \sum_{\stackrel{0 \le d \le L_y}{d=2 \ {\rm mod} \ 3}}
\sum_{j=1}^{n_P(L_y,d)} (\lambda_{P,L_y,d,j}(q=1))^m \ . 
\label{pq1}
\eeqs
However, in contrast to the case for $q=0$, the cancellation here is not
pairwise because some of the $\lambda$'s vanish.  Specifically, as will be
shown below, the term $\lambda_{P,L_y,d,j}$ for $d=L_y-1$ and $j=1$ is 
$q-1$, which vanishes at $q=1$.

\section{Sum Rules Applicable to M\"obius Strips of the Square Lattice}

\subsection{Relations Between Sums of $n_{Z,Mb}(L_y,d)$}

{\bf Prop. 9} \quad For the M\"obius strip of the square lattice of
width $L_y$ and arbitrary length $L_x$
\beqs
& &   \sum_{\stackrel{0 \le d \le d_{max,Mb}}{d \ {\rm even}}}
[n_{Z,Mb}(L_y,d,+) - n_{Z,Mb}(L_y,d,-)] \cr\cr
& = & \sum_{\stackrel{0 \le d \le d_{max,Mb}}{d \ {\rm odd}}}
[n_{Z,Mb}(L_y,d,+) - n_{Z,Mb}(L_y,d,-)] \ . 
\label{nzmbq0}
\eeqs

\medskip

{\bf Proof} \quad To prove this, we evaluate (\ref{czsummbcd}) for $q=0$ and 
use (\ref{cdq0}).  $\Box$

Applying the general fact that for any graph $G$,
$Z(G,q,v)=0$ for $q=0$, to the M\"obius strip of
the square lattice and using (\ref{cdq0}), we obtain
\beqs
& & Z(sq[Mb,L_y \times m],q=0,v) = 0 \cr\cr
& = & \sum_{\stackrel{0 \le d \le d_{max,Mb}}{\eta=\pm 1; \ C=+1}} 
                                        (\lambda_{Z,L,d,\eta,j}(q=0))^m - 
      \sum_{\stackrel{0 \le d \le d_{max,Mb}}{\eta=\pm 1; \ C=-1}} 
                                        (\lambda_{Z,L,d,\eta,j}(q=0))^m 
\cr\cr & & 
\label{zmbq0}
\eeqs
where 
\beq
C = \eta (-1)^d \ . 
\label{c}
\eeq
Since the cancellation between the various $m$'th powers of the eigenvalues
must occur for arbitrary $m$, they must, in fact, cancel in pairs.  This
provides another way of understanding the result in (\ref{nzmbq0}). 

\medskip

{\bf Prop. 10} \quad For the M\"obius strip of the square lattice of
width $L_y$ and arbitrary length $L_x$
\beqs
& &   \sum_{\stackrel{0 \le d \le d_{max,Mb}}{d=0 \ {\rm mod} \ 3}} 
[n_{Z,Mb}(L_y,d,+) - n_{Z,Mb}(L_y,d,-)] \cr\cr
& & +  \sum_{\stackrel{0 \le d \le d_{max,Mb}}{d=2 \ {\rm mod} \ 3}}
[-n_{Z,Mb}(L_y,d,+) + n_{Z,Mb}(L_y,d,-)] \cr\cr
& = & 1 \ . 
\label{nzmbq1}
\eeqs

\medskip

{\bf Proof} \quad To prove this, we evaluate (\ref{czsummbcd}) for $q=1$ and
use (\ref{cdq1}).  $\Box$

\medskip

{\bf Prop. 11} \quad For the M\"obius strip of the square lattice of
width $L_y$ and arbitrary length $L_x$
\beqs
& &   \sum_{\stackrel{0 \le d \le d_{max,Mb}}{d=0,1 \ {\rm mod} \ 4}}
[n_{Z,Mb}(L_y,d,+) - n_{Z,Mb}(L_y,d,-)] \cr\cr
& + & \sum_{\stackrel{0 \le d \le d_{max,Mb}}{d=2,3 \ {\rm mod} \ 4}}
[-n_{Z,Mb}(L_y,d,+) + n_{Z,Mb}(L_y,d,-)] \cr\cr
& = & 2^{[\frac{L_y+1}{2}]} \ .
\label{nzmbq2}
\eeqs

\medskip

{\bf Proof} \quad To prove this, we evaluate (\ref{czsummbcd}) for $q=2$ and
use (\ref{cdq2}).  $\Box$

\subsection{Relations Between Sums of $n_{P,Mb}(L_y,d)$}

{\bf Prop. 12} \quad For the M\"obius strip of the square lattice of width 
$L_y$ and arbitrary length $L_x$
\beqs
& &   \sum_{\stackrel{0 \le d \le d_{max,Mb}}{d \ {\rm even}}} 
[n_{P,Mb}(L_y,d,+) - n_{P,Mb}(L_y,d,-)] \cr\cr
& = & \sum_{\stackrel{0 \le d \le d_{max,Mb}}{d \ {\rm odd }}}  
[n_{P,Mb}(L_y,d,+) - n_{P,Mb}(L_y,d,-)] \ . 
\label{npmbq0}
\eeqs

\medskip

{\bf Proof} \quad To prove this, we evaluate (\ref{cpsummbcd}) for $q=0$ and
use (\ref{cdq0}).  $\Box$

Applying the general fact that for any graph $G$,
$P(G,q)=0$ for $q=0$, to the M\"obius strip of the
square lattice and using (\ref{cdq0}), we obtain the following relation, where
the $\lambda$'s are evaluated at $q=0$:
\beqs
& & P(sq[L_y \times m,Mb],q=0) = 0 \cr\cr
& = & \sum_{\stackrel{0 \le d \le d_{max,Mb}}{\eta=\pm 1; \ C=+1}}
        (\lambda_{P,L,d,\eta,j}(q=0))^m - 
      \sum_{\stackrel{0 \le d \le d_{max,Mb}}{\eta=\pm 1; \ C=-1}}
        (\lambda_{P,L,d,\eta,j}(q=0))^m \quad  \ . 
\label{pmbq0}
\eeqs
where $C$ was defined in (\ref{c}).  Since the cancellation between the various
$m$'th powers of the eigenvalues must occur for arbitrary $m$, they must, in
fact, cancel in pairs.  This is another way of understanding the result
(\ref{npmbq0}).

\medskip

{\bf Prop. 13} \quad For the M\"obius strip of the square lattice of width 
$L_y$ and arbitrary length $L_x$, 
\beqs
& &   \sum_{\stackrel{0 \le d \le d_{max,Mb}}{d=0 \ {\rm mod} \ 3}} 
[n_{P,Mb}(L_y,d,+) - n_{P,Mb}(L_y,d,-)] \cr\cr 
& = & \sum_{\stackrel{0 \le d \le d_{max,Mb}}{d=2 \ {\rm mod} \ 3}} 
[n_{P,Mb}(L_y,d,+) - n_{P,Mb}(L_y,d,-)] \ . 
\label{npmbq1}
\eeqs

\medskip

{\bf Proof} \quad To prove this, we evaluate (\ref{cpsummbcd}) for $q=1$ and
use (\ref{cdq1}).  $\Box$

\medskip

{\bf Prop. 14} \quad For the M\"obius strip of the square lattice of width
$L_y$ and arbitrary length $L_x$,
\beqs
& &   \sum_{\stackrel{0 \le d \le d_{max,Mb}}{d=0,1 \ {\rm mod} \ 4}}
[n_{P,Mb}(L_y,d,+) - n_{P,Mb}(L_y,d,-)] \cr\cr
& + & \sum_{\stackrel{0 \le d \le d_{max,Mb}}{d=2,3 \ {\rm mod} \ 4}}
[-n_{P,Mb}(L_y,d,+) + n_{P,Mb}(L_y,d,-)] \cr\cr
& = & 2 \ . 
\label{npmbq2}
\eeqs

\medskip

{\bf Proof} \quad To prove this, we evaluate (\ref{cpsummbcd}) for $q=2$ and
use (\ref{cdq2}).  $\Box$

\section{Sum Rules for Self-Dual Strips of the Square Lattice}

\subsection{Relations Between Sums of $n_{Z,G_D}(L_y,d)$}

We first note that for the self-dual strips of the square lattice $G_D(L_y
\times L_x)$, there is no analogue of the sum rules (\ref{nzq0}) and 
(\ref{npq0}) because all of the coefficients $\kappa^{(d)}$ have
$q$ as a factor and hence vanish at $q=0$.  We do, however, find several sum
rules.  

\medskip

{\bf Prop. 15 } \quad For the self-dual strip of the square lattice 
$G_D(L_y \times L_x)$
\beq
\sum_{\stackrel{1 \le d \le L_y+1}{d= 1 \ {\rm mod \ 3}}} n_{Z,G_D}(L_y,d) -
\sum_{\stackrel{1 \le d \le L_y+1}{d= 2 \ {\rm mod \ 3}}} n_{Z,G_D}(L_y,d) = 1
\ . 
\label{nzq1gd}
\eeq
\bigskip

{\bf Proof} \quad To prove this, we evaluate (\ref{czsumgd}) at $q=1$ and use 
(\ref{kappadq1}). $\Box$ 

\medskip

{\bf Prop. 16} \quad For the self-dual strip of the square lattice, 
$G_D(L_y \times L_x)$, 
\beq
\sum_{\stackrel{1 \le d \le L_y+1}{d=2k+1}} (-1)^k n_{Z,G_D}(L_y,d)=2^{L_y} \ .
\label{nzq2gd}
\eeq

\medskip

{\bf Proof} \quad To prove this, we evaluate (\ref{czsumgd}) at $q=2$ and use 
(\ref{kappadq2}). $\Box$ 

\medskip

\subsection{Relations Between Sums of $n_{P,G_D}(L_y,d)$ }

{\bf Prop. 17} \quad For the self-dual strip of the square lattice 
$G_D(L_y \times L_x)$, 
\beq
 \sum_{\stackrel{1 \le d \le L_y+1}{d=1 \ {\rm mod} \ 3}} n_{P,G_D}(L_y,d)
=\sum_{\stackrel{1 \le d \le L_y+1}{d=2 \ {\rm mod} \ 3}} n_{P,G_D}(L_y,d) \ .
\label{npq1gd}
\eeq

\medskip

{\bf Proof} \quad To prove this, we evaluate (\ref{cpsumgd}) at $q=1$ and use
(\ref{kappadq1}).  $\Box$ 

\medskip

{\bf Prop. 18} \quad For the self-dual strip of the square lattice 
$G_D(L_y \times L_x)$, 
\beq
\sum_{\stackrel{1 \le d \le L_y+1}{d=2k+1}} (-1)^k n_{P,G_D}(L_y,d)=1 \ .
\label{npq2gd}
\eeq

\medskip

{\bf Proof} \quad To prove this, we evaluate (\ref{cpsumgd}) at $q=2$ and use
(\ref{kappadq2}). $\Box$

As before for other strip graphs, one may analyze how cancellations occur 
in the chromatic polynomial at values of $q$ where it vanishes.  Thus, let us
consider the value $q=1$ and use the fact that for any connected graph
$G$ with at least two vertices, $P(G,q)=0$ if $q=1$.  Applying this to
$G=G_D(L_y \times L_x)$ and using the property (\ref{kappadq1}), we obtain the
following relation, where the $\lambda$'s are evaluated at $q=1$: 
\beqs
& & P(G_D(L_y \times m,q=1) = 0 \cr\cr
& = & \sum_{\stackrel{1 \le d \le L_y+1}{d=1 \ {\rm mod} \ 3}}
\sum_{j=1}^{n_{P,G_D}(L_y,d)} (\lambda_{P,G_D,L_y,d,j})^m 
 - \sum_{\stackrel{1 \le d \le L_y+1}{d=2 \ {\rm mod} \ 3}}
\sum_{j=1}^{n_{P,G_D}(L_y,d)} (\lambda_{P,G_D,L_y,d,j})^m \ . 
\label{pq1gd}
\eeqs
For the cases where we have carried out explicit calculations, we find that the
$\lambda$'s do not vanish at $q=1$ for this family of graphs.  Since the
cancellation in (\ref{pq1gd}) between the various $m$'th powers of the
eigenvalues must occur for arbitrary $m$, they must, in fact, cancel in pairs.

Since the family $G_D(L_y \times L_x)$ contains triangles, $P(G_D[L_y \times
L_x],q)=0$ for $q=2$.  One may investigate what is implied by this property.
Using (\ref{kappadq2}), one finds 
\beqs
& & P(G_D[L_y \times m],q=2) = 0 \cr\cr
& = & 2\sum_{\stackrel{1 \le d \le L_y+1}{d=2k+1}} (-1)^k 
\sum_{j=1}^{n_P(G_D,L_y,d)} (\lambda_{P,G_D,L_y,d,j}(q=2))^m \ . 
\label{pq2gd}
\eeqs
However, this cancellation does not, in general, occur in a pairwise manner,
because some of the $\lambda_{P,G_D,L_y,2k+1,j}$ vanish.  For example, for
$L_y=2$, $n_P(G_D,1)=2$ while $n_P(G_D,3)=1$, but the $\lambda_{P,G_D,2,1,2}$,
given by eq. (6.3) in \cite{dg},
\beq
\lambda_{P,G_D,2,1,j}= \frac{1}{2}\bigg [q^2-5q+7 \pm (q^4-6q^3+15q^2-22q+17
)^{1/2} \bigg ] \quad j=1,2
\label{lamgd2d1j}
\eeq
vanishes at $q=2$.

\section{Sum Rules for Other Recursive Families of Graphs}

Proceeding in a manner similar to that above, one can obtain analogous sum
rules for other recursive families of graphs.  We comment briefly on one such
family.  In \cite{ka3} we presented theorems determining the structure of the
Potts model partition function for a cyclic clan graph (CG).  We recall some
relevant definitions.  A complete graph $K_r$ is a graph containing $r$
vertices with the property that each vertex is connected by edges to every
other vertex.  The join of two graphs $H_1$ and $H_2$, denoted $H_1+H_2$, is
the graph formed by connecting each vertex of $H_1$ to all of the vertices of
$H_2$ with edges.  Then a (homogeneous) cyclic clan graph, denoted
$G[(K_r)_m,jn]$ is a recursive graph composed of a set of $m$ complete graphs
$K_r$, such that the linkage between two adjacent pairs of $K_r$'s is a join
(abbreviated $jn$).  The Potts model partition function has the form \cite{ka3}
\beq
Z(G[(K_r)_m,jn],q,v) = \sum_{d=0}^r \mu_d \sum_{j=1}^{n_{Z,CG}(r,d)}
(\lambda_{Z,CG,r,d,j})^m 
\label{zgsumclan}
\eeq
where $\mu_0=1$ and 
\beq
\mu_d = {q \choose d} - {q \choose d-1} = \frac{q_{(d-1)}(q-2d+1)}{d!}
\quad {\rm for} \quad 1 \le d \le r 
\label{mud}
\eeq
where
\beq
q_{(r)}=\prod_{s=0}^{r-1}(q-s) 
\label{ff}
\eeq
is the falling factorial in combinatorics.  The numbers $n_{Z,CG}(r,d)$ were
determined in \cite{ka3}.  

Let us apply the relation $Z(G,q=0,v)=0$ to this family. All of the $\mu_d$
coefficients for $d \ge 2$ contain $q$ as a factor and hence vanish identically
for $q=0$.  Using $\mu_1 = q-1$, the condition $Z(G,q=0,v)=0$ reduces to
\beq
\sum_{j=1}^{n_{Z,CG}(r,0)} (\lambda_{Z,CG,r,0,j})^m = 
\sum_{j=1}^{n_{Z,CG}(r,1)} (\lambda_{Z,CG,r,1,j})^m \ . 
\label{clansum}
\eeq
Since this must hold for arbitrary $m$ and since, in general, the relevant 
$\lambda$'s are nonvanishing, it follows that the cancellation must occur in a
pairwise manner, and hence that $n_{Z,CG}(r,0)=n_{Z,CG}(r,1)$.  This is in
agreement with a result that we already derived in \cite{ka3}, viz., that 
$n_{Z,CG}(r,0)=n_{Z,CG}(r,0)=2^{r-1}$.  So here one does not obtain any new sum
rule.  A similar comment applies for the chromatic polynomial
$P(G[(K_r)_m,jn],q)$, where $n_{P,CG}(r,d)=1$ for all $d$ (with $0 \le d \le
r$) \cite{readcarib81}. 

\section{Determination of $\lambda_{P,L_y,d=L_y-1}$ for Families of Lattice 
Strip Graphs}
\label{lamlyminus1}

\subsection{Cyclic and M\"obius Strips of the Square Lattice} 

We have succeeded in determining the terms $\lambda_{P,L_y,d=L_y-1}$ with
coefficient $c^{(L_y-1)}$ in the chromatic polynomial for the cyclic and
M\"obius strips of the square lattice.  There are $n_P(L_y,L_y-1)=L_y$ of these
terms, by eq. (\ref{nplym1}).  Since all of the results in this section and the
rest of the paper (including the appendix) refer to the chromatic polynomial,
we shall use an abbreviated notation omitting the subscript $P$ in the terms
$\lambda_{P,L_y,d}$.  For our calculation, we use the sieve method of
\cite{matmeth} to calculate the $L_y \times L_y$ transfer matrix for these
$\lambda$'s with coefficient $c^{(L_y-1)}$.  This transfer matrix is denoted
$T^{(L_y,d)}$.  To avoid awkward notation, we leave the type of lattice
implicit.  In subsequent sections, when the same symbol is used for strips of
other lattices, it will be understood that its meaning is specific to those
sections.  It is convenient to extract a prefactor and define a reduced
transfer matrix:
\beq
T^{(L_y,L_y-1)} = (-1)^{L_y+1}\bar T^{(L_y,L_y-1)} \ . 
\label{tbardef}
\eeq

\medskip

{\bf Theorem 1} \quad For the cyclic or M\"obius strip of the square lattice
with arbitrary width $L_y$ and arbitrary length $L_x=m$, the eigenvalues 
$\lambda_{L_y,d,j}$ of for coefficient degree $d=L_y-1$ are
\beq 
\lambda_{L_y,d=L_y-1,j} = (-1)^{L_y+1}(q-a_{sq,L_y,j}) \quad , \ 1 \le j
\le L_y
\label{lamdlym1}
\eeq
where
\beq
a_{sq,L_y,j} = 1 + 4\cos^2 \left ( \frac{(L_y+1-j) \pi}{2L_y} \right )
\quad , \ 1 \le j \le L_y \ . 
\label{asqlyj}
\eeq

\medskip

{\bf Proof} \quad 

We begin with the following lemma: 

\medskip
 
{\bf Lemma 1} \quad For the cyclic strip of the square lattice with arbitrary
width $L_y$ and length $L_x=m$, denoted $sq[L_y \times L_x,,cyc.]$, the matrix
$\bar T^{(L_y,L_y-1)}$ for $L_y \ge 2$ is given by
\beq
\bar T^{(L_y,L_y-1)}_{11} = \bar T^{(L_y,L_y-1)}_{L_y \ L_y} = q-2
\label{tbarqm1}
\eeq

\beq
\bar T^{(L_y,L_y-1)}_{jj} = q-3 \quad {\rm for} \quad 2 \le j \le L_y-1
\label{tbarqm3}
\eeq

\beq
\bar T^{(L_y,L_y-1)}_{j \ j+1} = \bar T^{(L_y,L_y-1)}_{j+1 \ j} = -1 \quad 
{\rm for} \quad 1 \le j \le L_y-1 
\label{tbardiags}
\eeq
with other elements equal to zero. For $L_y=1$, $\bar
T^{(1,0)}=\lambda_{1,0}=q-1$.

\medskip

Thus, for example, for $L_y=5$,
\beq
\bar T^{(5,4)}=\left(
  \begin{array}{ccccc}
   q-2 & -1  &  0  &  0  &  0  \cr
   -1  & q-3 & -1  &  0  &  0  \cr
    0  & -1  & q-3 & -1  &  0  \cr
    0  & 0   & -1  & q-3 & -1  \cr
    0  & 0   &  0  & -1  & q-2 
    \end{array}\right ) \ .
\label{tbar54}
\eeq
We note that $\bar T^{(L_y,L_y-1)}$ is a special case of a Jacobi matrix, in
the terminology of linear algebra (where a Jacobi matrix $A$ is defined as a
matrix with the property that $A_{ij}=0$ if $|i-j| \ge 2$) \cite{marcus}. 

\medskip

{\bf Proof} \quad One can construct the cyclic strip of the square lattice via
the repetition of a subgraph which is a transverse slice consisting of the path
graph with $L_y$ vertices such that the linkage $L$ between two successive such
path graphs is the identity linkage, so that vertex 1 of the first path graph
is connected by an edge to vertex 1 of the next path graph, and so forth for
all $L_y$ vertices.  We denote this linkage as $L =
\{(1,1),(2,2),...,(L_y,L_y)\}$. Now applying the sieve method of
\cite{matmeth}, we consider the functions $\alpha : V \rightarrow
\{1,2,...,q\}$ that are proper $q$-coloring of the path (or tree) graph with
$L_y$ vertices, so that $V$ is the vertex set $V = \{1,2,...,L_y\}$. Let us
choose the basis so that
\beq
[\alpha](\beta) = \cases{ 1 & if $\alpha = \beta$ \cr
                          0 & otherwise } \ .
\label{alphabeta}
\eeq
Define the compatibility matrix as follows,
\beq (T)_{\alpha\beta} = \cases{ 1 & if $(\alpha,\beta)$ is compatible
with $L$ \cr 0 & otherwise } \ .
\label{talphabeta}
\eeq

Denote $[1|h]$ as the function that takes the value 1 for the coloring that
assign color $h$ to vertex 1, and takes the value 0 otherwise. Similarly let
$[1,2,...|h_1,h_2,...]$ refer to the coloring choice such that color $h_1$
is assigned to vertex 1, color $h_2$ is assigned to vertex 2 and so forth. The
coloring without restriction is denoted $u$. We have
\beq
T[\alpha] = \sum _{X\subseteq V} (-1)^{|X|}[X|\alpha]
\label{talpha}
\eeq
where $X$ refers to all subsets of the $L_y$ vertices of the path graph, and 
$|X|$ is the number of vertices in $X$. For example, for $L_y=3$, we have 
\beq
T[\alpha] = u - [1|\alpha_1] - [2|\alpha_2] - [3|\alpha_3] +
[1,2|\alpha_1,\alpha_2] + [1,3|\alpha_1,\alpha_3] + 
[2,3|\alpha_2,\alpha_3]  -  [1,2,3|\alpha_1,\alpha_2,\alpha_3] \ .
\label{talphaLy3}
\eeq
Concentrating now on the case $d = L_y-1$, we observe that the invariant
subspace of the matrix $T$ is spanned by
\beq
[1,2,...,j-1,j+1,...,L_y|h_1,h_2,...,h_{j-1},h_{j+1},...,h_{L_y}], \quad
1\le j \le L_y \ . 
\label{subspace}
\eeq
Therefore,
\beq
T[1,2,...,j-1,j+1,...,L_y|h_1,h_2,...,h_{j-1},h_{j+1},...,h_{L_y}] =
T \left ( \sum_{H} [\alpha] \right ) = \sum_{H} T[\alpha]
\label{th}
\eeq
where $H$ means $\alpha_1=h_1, \alpha_2=h_2,...,\alpha_{j-1}=h_{j-1},
\alpha_{j+1}=h_{j+1},...,\alpha_{L_y}=h_{L_y}$, and \newline
$h_1, h_2,...,h_{j-1},h_{j+1},...,h_{L_y}$ 
are different from each other.

To obtain the matrix for level $L_y-1$, we only have to consider the 
terms in the summation of eq. (\ref{talpha}) with $|X|=L_y-1$ and 
$|X|=L_y$. Now the first row of $T^{(L_y,L_y-1)}$ is obtained by applying
$T$ on $[1,2,...,L_y-1|h_1,h_2,...,h_{L_y-1}]$. The nonzero elements are
\beqs
& & (-1)^{L_y-1}((q-1)[1,2,...,L_y-1|h_1,h_2,...,h_{L_y-1}] +
(-[1,2,...,L_y-2,L_y|h_1,h_2,...,h_{L_y-2},h_{L_y}])) \cr\cr
& & + (-1)^{L_y}[1,2,...,L_y-1|h_1,h_2,...,h_{L_y-1}] \cr\cr
& = & (-1)^{L_y-1}((q-2)[1,2,...,L_y-1|h_1,h_2,...,h_{L_y-1}] \cr\cr
& - & [1,2,...,L_y-2,L_y|h_1,h_2,...,h_{L_y-2},h_{L_y}])
\label{firstrow}
\eeqs
and we have the corresponding result for the last row. Applying $T$ on 
$[1,2,...,j-1,j+1,...,L_y|h_1,h_2,...,h_{j-1},h_{j+1},...,h_{L_y}]$ for
$2 \le j \le L_y-1$, we find that the nonzero elements are
\beqs
& & 
(-1)^{L_y-1}(-[1,2,...,j,j+2,...,L_y|h_1,h_2,...,h_{j},h_{j+2},...,h_{L_y}]
\cr\cr
& & + 
(q-2)[1,2,...,j-1,j+1,...,L_y|h_1,h_2,...,h_{j-1},h_{j+1},...,h_{L_y}]
\cr\cr 
& & + 
(-[1,2,...,j-2,j,...,L_y|h_1,h_2,...,h_{j-2},h_{j},...,h_{L_y}])) \cr\cr 
& & + 
(-1)^{L_y}[1,2,...,j-1,j+1,...,L_y|h_1,h_2,...,h_{j-1},h_{j+1},...,h_{L_y}]
\cr\cr
& = &
(-1)^{L_y-1}(-[1,2,...,j,j+2,...,L_y|h_1,h_2,...,h_{j},h_{j+2},...,h_{L_y}]
\cr\cr
& & + 
(q-3)[1,2,...,j-1,j+1,...,L_y|h_1,h_2,...,h_{j-1},h_{j+1},...,h_{L_y}]
\cr\cr 
& & - [1,2,...,j-2,j,...,L_y|h_1,h_2,...,h_{j-2},h_{j},...,h_{L_y}])
\label{otherrow}
\eeqs
and, taking into account the relation (\ref{tbardef}), the lemma follows for
the case of the cyclic strip. It was proved in \cite{tor4} that the $\lambda$'s
for the cyclic and M\"obius strips of the square lattice are identical to each
other, and similarly for the triangular lattice.  Finally, for the $L_y=1$
case, $T^{(1,0)}=\lambda_{1,0}=q-1$ is obtained by an explicit elementary
calculation.  This completes the proof of the lemma. \ $\Box$

\medskip

Next, we have 

\medskip

{\bf Lemma 2} \quad  The eigenvalues of the (reduced) transfer matrix 
$\bar T^{(L_y,L_y-1)}$ are $q-a_{L_y,j}$ for $1 \le j \le L_y$.  

\medskip

{\bf Proof} \quad For the case $L_y=1$, an elementary explicit calculation
yields the result.  For $L_y \ge 2$, consider the $L_y \times L_y$ matrix
$A(L_y)$ given by
\beq
A(L_y)_{j \ j+1} = A(L_y)_{j+1 \ j} = -1 \quad
{\rm for} \quad 1 \le j \le L_y-1
\label{adiags}
\eeq
with all other elements equal to zero, and $A(L_y)=0$ for $L_y=1$. Denote the
characteristic polynomial of $A(L_y)$ in terms of variable $\omega$ as
$C[A(L_y)]={\rm det}(\omega I - A(L_y))$, where $I$ is the $L_y \times
L_y$ identity matrix.  This polynomial satisfies the recursion relation (for
$L_y \ge 2$) 
\beq
C[A(L_y)] = \omega C[A(L_y-1)] - C[A(L_y-2)]
\label{arecursion}
\eeq
where, for $L_y=2$ we formally set $C[A(0)] \equiv 1$.  As in our earlier
related work \cite{cf,dg}, we observe that this is the same as the recursion
relation satisfied by the Chebyshev polynomial of the second kind, $U_n(x)$, 
for $x=\omega/2$, i.e.,
\beq
U_n(x) = 2x U_{n-1}(x)-U_{n-2}(x) \ .
\label{unrecursion}
\eeq
Solving the recursion relation (\ref{arecursion}), we obtain 
\beq
C[A(L_y)] = U_{L_y}(\omega /2) = \sum_{j=0}^{[\frac{L_y}{2}]} (-1)^j {L_y-j
\choose j}(\omega)^{L_y-2j} 
\label{achar}
\eeq
where we again use the notation $[\nu]$ to denote the integral part of $\nu$.
Using the relation 
\beq
U_n(\cos \phi) = \frac{\sin((n+1)\phi)}{\sin \phi}
\label{unsin}
\eeq
we express this equivalently as 
\beq
C[A(L_y)]=\frac{\sin[(L_y+1){\rm arccos}(\omega /2)]}
{\sin[{\rm arccos}(\omega /2)]}
\label{achar2}
\eeq
It follows that the roots of the charactistic polynomial $C[A(L_y)]$, i.e., the
eigenvalues of the matrix $A(L_y)$, are
\beq
\omega_j = 2\cos(\frac{j\pi}{L_y+1}) \quad {\rm for} \ 1 \le j \le L_y \ .
\label{acharsol}
\eeq
Now the reduced transfer matrix $\bar T^{(L_y,L_y-1)}$ can be written for $L_y
\ge 2$ as 
\beq
\bar T^{(L_y,L_y-1)} = (q-3) I + \tilde T^{(L_y,L_y-1)}
\label{sqtbarttilde}
\eeq
and the matrix $\tilde T^{(L_y,L_y-1)}$ is given by 
\beq
\tilde T^{(L_y,L_y-1)}_{11} = \tilde T^{(L_y,L_y-1)}_{L_y \ L_y} = 1
\label{sqttilde}
\eeq
\beq
\tilde T^{(L_y,L_y-1)}_{j \ j+1} = \tilde T^{(L_y,L_y-1)}_{j+1 \ j} = -1
\quad {\rm for} \quad 1 \le j \le L_y-1
\label{sqttildediags}
\eeq
with all other elements 0. Let us denote the characteristic polynomial of
$\tilde T^{(L_y,L_y-1)}$ in terms of the variable $\omega$ as $C[\tilde
T^{(L_y,L_y-1)}]$.  We find that this characteristic polynomial satisfies the
following recursion relation 
\beq
C[\tilde T^{(L_y,L_y-1)}]=(\omega -2) C[A(L_y-1)] \ .
\label{ttildechar}
\eeq
Using eq. (\ref{acharsol}), we find that the roots of $C[\tilde
T^{(L_y,L_y-1)}]$ are given by
\beq
\omega_j = 2\cos \biggl ( \frac{j\pi}{L_y} \biggr ) \quad {\rm for} 
\ 0 \le j \le L_y-1 \ .
\label{ttildacharsol}
\eeq
Combining eqs. (\ref{sqtbarttilde}) and ({\ref{ttildacharsol}) (and relabelling
the roots so that the $a_{L_y,j}$ increase monotonically as a function of $j$)
then yields the result in the lemma. $\Box$

\medskip

Finally, we use the result from \cite{pm,tor4} that the $\lambda_{P,G,j}$'s for
the cyclic and M\"obius strips graphs $G$ of a given lattice and width $L_y$
are the same.  The theorem then follows. $\Box$ 

\medskip

We remark that eqs. (5.2) and (5.3) in \cite{s4} are in accord with this
structure of the transfer matrix.

\medskip

Several corollaries of Theorem 1 and the associated lemmas are of
interest. Each of these applies to the cyclic or M\"obius strips of the square
lattice; we leave this implicit in the notation.  Since the proofs are
straightforward, we omit them.

\medskip

{\bf Corollary 1} \quad  The trace and determinant of $\bar T^{(L_y,L_y-1)}$ 
are
\beq
{\rm Tr}(\bar T^{(L_y,L_y-1)}) = 2+L_y(q-3)
\label{lamsqtrace}
\eeq
and
\beq
{\rm det}(\bar T^{(L_y,L_y-1)}) = \prod_{j=1}^{L_y}(q-a_{sq,L_y,j}) 
\label{lamsqprod}
\eeq
where the $a_{sq,L_y,j}$ were given in (\ref{asqlyj}). 

\medskip

{\bf Corollary 2} \quad 

For arbitrary $L_y$, $q-1$ is an eigenvalue of $\bar T^{(L_y,L_y-1)}$ 
with multiplicity 1, given by 
\beq
\lambda_{L_y,j=1}=q-1 \ . 
\label{lamqm1}
\eeq

\medskip

{\bf Corollary 3} \quad 

If $L_y = 0$ mod 2, i.e., $L_y \ge 2$ is even, then $q-3$ is an eigenvalue 
of $\bar T^{(L_y,L_y-1)}$ with multiplicity 1 given by 
\beq
\lambda_{L_y,j=1+(L_y/2)}=q-3 \ . 
\label{lamqm3}
\eeq

\medskip

{\bf Corollary 4} 

If $L_y=0$ mod 3, then $q-2$ and $q-4$ are eigenvalues of 
$\bar T^{(L_y,L_y-1)}$, each with multiplicity
1, given by
\beq
\lambda_{L_y,j=1+(L_y/3)}=q-2
\label{lamqm2}
\eeq
\beq
\lambda_{L_y,j=1+(2L_y/3)}=q-4 \ . 
\label{lamqm4}
\eeq

{\bf Corollary 5} 

The roots $a_{sq,L_y,j}$ monotonically increases from 1, for $j=1$, to
$1+4\cos^2(\pi/(2L_y))$ for $j=L_y$.  As $L_y \to \infty$, these roots
become dense on the interval $[1,5]$. 

\bigskip

\subsection{Self-Dual Cyclic Strips of the Square Lattice}

{\bf Theorem 2} \quad For the self-dual cyclic strip of the square lattice
with arbitrarily width $L_y$ and length $L_x=m$, the eigenvalues 
$\lambda_{P,L_y,d,j}$ for coefficient degree $d=L_y-1$ are
\beq 
\lambda_{L_y,d=L_y-1,j} = (-1)^{L_y+1}(q-a_{G_D,L_y,j}) \quad , \ 1 \le j
\le L_y 
\label{lamdbc2dlym1}
\eeq
where
\beq
a_{G_D,L_y,j} = 1 + 4\cos^2 \left ( \frac{(2j-1) \pi}{2L_y+1} \right )
\quad , \ 1 \le j \le L_y \ .
\label{adbc2lyj}
\eeq

\medskip

{\bf Proof} \quad  We begin with the following lemma: 

\medskip
 
{\bf Lemma 3} \quad For the self-dual cyclic strip of the square lattice with
arbitrarily width $L_y$ and length $L_x=m$, the reduced transfer matrix 
$\bar T^{(L_y,L_y-1)}$ for $L_y \ge 2$ is given by
\beq
\bar T^{(L_y,L_y-1)}_{11} = q-2
\label{tbarsdqm1}
\eeq

\beq
\bar T^{(L_y,L_y-1)}_{jj} = q-3 \quad {\rm for} \quad 2 \le j \le L_y
\label{tbarsdqm3}
\eeq

\beq
\bar T^{(L_y,L_y-1)}_{j \ j+1} = \bar T^{(L_y,L_y-1)}_{j+1 \ j} = -1 \quad 
{\rm for} \quad 1 \le j \le L_y-1 
\label{tbarsddiags}
\eeq
with other elements equal to zero. For $L_y=1$, $\bar T^{(1,0)}=\lambda_{1,0}=q-2$.

\medskip

{\bf Proof} \quad This proceeds in a manner similar to the proof given above,
with the difference that the coloring is further constrained by the feature
that all of the vertices on the upper side of the strip are connected by edges
to a single external vertex.  This difference has the effect of replacing the
$q-2$ by $q-3$ in the entry $\bar T^{(L_y,L_y-1)}_{L_y L_y}$.  $\Box$ 

\medskip

An example for $L_y=5$ is 

\beq
\bar T^{(5,4)}=\left(
  \begin{array}{ccccc}
   q-2 & -1  &  0  &  0  &  0  \cr
   -1  & q-3 & -1  &  0  &  0  \cr
    0  & -1  & q-3 & -1  &  0  \cr
    0  & 0   & -1  & q-3 & -1  \cr
    0  & 0   &  0  & -1  & q-3
    \end{array}\right ) \ .
\label{tbardbc254}
\eeq

\medskip

Next, 

\medskip

{\bf Lemma 4} \quad The eigenvalues of $\bar T^{(L_y,L_y-1)}$ are 
given by 
\beq
q-a_{G_D,L_y,j} \quad {\rm for} \ \  1 \le j \le L_y \ .
\label{lambdasd}
\eeq

\medskip

{\bf Proof} \quad For the case $L_y=1$, an explicit calculation yields the
result.  For $L_y \ge 2$, consider the $L_y \times L_y$ matrix $A(L_y)$ given
by
\beq
\bar T^{(L_y,L_y-1)} = (q-3) I + \tilde T^{(L_y,L_y-1)}
\label{gdtbarttilde}
\eeq
where again $I$ is the $L_y \times L_y$ identity matrix, and the matrix $\tilde
T^{(L_y,L_y-1)}$ is given by 
\beq
\tilde T^{(L_y,L_y-1)}_{11} = 1
\label{gdttilde}
\eeq

\beq
\tilde T^{(L_y,L_y-1)}_{j \ j+1} = \tilde T^{(L_y,L_y-1)}_{j+1 \ j} = -1
\quad {\rm for} \quad 1 \le j \le L_y-1
\label{gdttildediags}
\eeq
with all other elements 0. Let us denote the characteristic polynomial of
$\tilde T^{(L_y,L_y-1)}$ in terms of the variable $\omega$ as $C[\tilde
T^{(L_y,L_y-1)}]$.  We observe that this characteristic polynomial satisfies
the recursion relations for $L_y \ge 2$ 
\beqs
C[\tilde T^{(L_y,L_y-1)}] & = & (\omega -1)C[A(L_y-1)] - C[A(L_y-2)] \cr\cr
                          & = & C[A(L_y)] - C[A(L_y-1)] 
\label{gdttildechar}
\eeqs
where the matrix $A(L_y)$ and its characteristic polynomial $C[A(L_y)]$ were
given in Lemma 2, and we continue to use the formal definition $C[A(0)]=1$.
Using eq. (\ref{achar2}), we find that $C[\tilde T^{(L_y,L_y-1)}]$ can be
written as
\beqs
C[\tilde T^{(L_y,L_y-1)}] & = & \frac{\sin[(L_y+1){\rm arccos}(\omega/2)]} 
{\sin[{\rm arccos}(\omega/2)]} - \frac{\sin[L_y {\rm arccos}(\omega/2)]} 
{\sin[{\rm arccos}(\omega/2)]} \cr\cr
& = & \frac{2\cos[(L_y+1/2){\rm arccos}(\omega/2)] 
\sin[{\rm arccos}(\omega/2)/2]} {\sin[{\rm arccos}(\omega/2)]} \cr\cr
& = & \frac{\cos[(L_y+1/2){\rm arccos}(\omega/2)]} 
{\cos[{\rm arccos}(\omega/2)/2]} \ .
\label{ttildechar2}
\eeqs
Therefore, the roots of the characteristic polynomial 
$C[\tilde T^{(L_y,L_y-1)}]$ are given by
\beq
\omega_j = 2\cos \biggl ( \frac{(2j-1)\pi}{2L_y+1} \biggr ) \quad {\rm for} 
\ 1 \le j \le L_y \ . 
\label{gdttildacharsol}
\eeq
Combining eqs. (\ref{gdtbarttilde}) and ({\ref{gdttildacharsol}) then yields
the result in the lemma. $\Box$ 

\medskip

Finally, combining Lemmas 3 and 4 yields the theorem. $\Box$ 

As is the case with $a_{sq,L_y,j}$, the roots $a_{G_D,L_y,j}$ become dense on
the interval $[1,5]$ as $L_y \to \infty$.  It is also straightforward to derive
the following corollary:

\medskip

{\bf Corollary 6} \quad If $L_y=1$ mod 3, then $q-2$ is a root of the reduced
transfer matrix $\bar T^{(L_y,L_y-1)}$ for the self-dual cyclic strip of the
square lattice of width $L_y$.

\medskip 

\medskip

\subsection{Cyclic and M\"obius Strips of the Triangular Lattice}

We have also determined the corresponding $L_y \times L_y$ transfer matrix for
the triangular lattice.  As noted before, to avoid awkward notation, we shall
use the same symbol for the transfer matrix $T^{(L_y,d)}$, but understand that
it is different for this lattice strip than for the previous strips. 
We have the following theorem for $\bar T^{(L_y,L_y-1)}$,

\medskip

{\bf Theorem 3} \quad The general matrix $\bar T^{(L_y,L_y-1)}$ for the cyclic
strip of the triangular lattice with $L_y \ge 2$ is given by
\beq
\bar T^{(L_y,L_y-1)}_{11} = q-3
\label{tbartriqm3}
\eeq
\beq
\bar T^{(L_y,L_y-1)}_{i \ j} = q-4 \quad {\rm for} \quad 2 \le i
\le
L_y-1, \quad 1 \le j \le i
\label{tbartriqm4}
\eeq
\beq
\bar T^{(L_y,L_y-1)}_{L_y \ j} = q-2 \quad {\rm for} \quad 1 \le j  
\le L_y
\label{tbartriqm2}
\eeq
\beq
\bar T^{(L_y,L_y-1)}_{j \ j+1} = -1 \quad {\rm for} \quad 1 \le j
\le L_y-1 \ . 
\label{tbartrim1}
\eeq

\medskip

{\bf Proof} \quad We use the sieve formula again.  Consider the strip of the
triangular lattice of width $L_y$ and arbitrary length $L_x$ to be constructed
by starting with the corresponding cyclic strip of the square lattice and
adding edges connecting the upper right and lower left vertices of the squares
on the strip, so that the edge set is
$L=\{(1,1),(2,1),(2,2),(3,2),...(L_y,L_y)\}$. In this case, there is more than
one edge connecting a vertex on one path graph forming a transverse slice to
the adjacent path graph, so eq. (\ref{talpha}) has to be written in a more
general form,
\beq
T[\alpha] = \sum _{X\subseteq V} (-1)^{|X|} \sum _\ell [X|\alpha_\ell]
\label{talphagen}
\eeq
where we sum over all different sets of edges connecting the vertices of the
right-hand path graph, $X$, to the vertices of the left-hand path graph for
each such pair. As an example, for the $L_y=3$ case, this is
\beqs
T[\alpha] & = & u - [1|\alpha_1] - [1|\alpha_2] - [2|\alpha_2] -
[2|\alpha_3] - [3|\alpha_3] + [1,2|\alpha_1,\alpha_2] +
[1,2|\alpha_1,\alpha_3] + [1,2|\alpha_2,\alpha_3] \cr\cr
& & + [1,3|\alpha_1,\alpha_3] + [1,3|\alpha_2,\alpha_3] +
[2,3|\alpha_2,\alpha_3] - [1,2,3|\alpha_1,\alpha_2,\alpha_3] \ . 
\label{talphatriLy3}
\eeqs
For $d = L_y-1$, the invariant subspace of $T$ is again spanned by
the terms given in eq. (\ref{subspace}). The nonzero elements in the
first row of $T^{(L_y,L_y-1)}$ are 
\beqs
& & (-1)^{L_y-1}((q-1)[1,2,...,L_y-1|h_1,h_2,...,h_{L_y-1}] +
(-[1,2,...,L_y-1|h_1,h_2,...,h_{L_y-1}] ) \cr\cr
& & + (-[1,2,...,L_y-2,L_y|h_1,h_2,...,h_{L_y-2},h_{L_y}])) +
(-1)^{L_y}[1,2,...,L_y-1|h_1,h_2,...,h_{L_y-1}] \cr\cr
& = & (-1)^{L_y-1}((q-3)[1,2,...,L_y-1|h_1,h_2,...,h_{L_y-1}] -
[1,2,...,L_y-1|h_1,h_2,...,h_{L_y-2},h_{L_y}]) \ . \cr\cr
& &
\label{firstrowtri}
\eeqs
For row $j$ from $2$ to $L_y-1$, the nonzero elements are
\beqs
& & (-1)^{L_y-1}(-[1,2,...,L_y-1|h_1,h_2,...,h_{L_y-1}] + 
(q-2)[1,2,...,L_y-1|h_1,h_2,...,h_{L_y-1}] \cr\cr
& & + (-[1,2,...,L_y-1|h_1,h_2,...,h_{L_y-1}]) + ... \cr\cr
& & 
+ (-[1,2,...,j,j+2,...,L_y|h_1,h_2,...,h_{j},h_{j+2},...,h_{L_y}])
\cr\cr
& & + (q-2)[1,2,...,j,j+2,...,L_y|h_1,h_2,...,h_{j},h_{j+2},...,h_{L_y}]
\cr\cr
& & +
(-[1,2,...,j,j+2,...,L_y|h_1,h_2,...,h_{j},h_{j+2},...,h_{L_y}]) \cr\cr
& & +
(q-2)[1,2,...,j-1,j+1,...,L_y|h_1,h_2,...,h_{j-1},h_{j+1},...,h_{L_y}]
\cr\cr
& & +
(-[1,2,...,j-1,j+1,...,L_y|h_1,h_2,...,h_{j-1},h_{j+1},...,h_{L_y}]) \cr\cr
& & +
(-[1,2,...,j-2,j,...,L_y|h_1,h_2,...,h_{j-2},h_{j},...,h_{L_y}])) \cr\cr
& & + 
(-1)^{L_y}[1,2,...,j-1,j+1,...,L_y|h_1,h_2,...,h_{j-1},h_{j+1},...,h_{L_y}]
\cr\cr
& = &
(-1)^{L_y-1}((q-4)[1,2,...,L_y-1|h_1,h_2,...,h_{L_y-1}] + ... \cr\cr
& & + (q-4)[1,2,...,j,j+2,...,L_y|h_1,h_2,...,h_{j},h_{j+2},...,h_{L_y}] 
\cr\cr
& & + 
(q-4)[1,2,...,j-1,j+1,...,L_y|h_1,h_2,...,h_{j-1},h_{j+1},...,h_{L_y}]
\cr\cr 
& & - [1,2,...,j-2,j,...,L_y|h_1,h_2,...,h_{j-2},h_{j},...,h_{L_y}]) \ . 
\label{otherrowtri}
\eeqs
The nonzero elements for the last row are
\beqs
& & (-1)^{L_y-1}(-[1,2,...,L_y-1|h_1,h_2,...,h_{L_y-1}] + 
(q-1)[1,2,...,L_y-1|h_1,h_2,...,h_{L_y-1}] + ... \cr\cr
& & + (-[1,3,...,L_y|h_1,h_3,...,h_{L_y}]) +   
(q-1)[1,3,...,L_y|h_1,h_3,...,h_{L_y}] \cr\cr
& & + (q-1)[2,3,...,L_y|h_2,h_3,...,h_{L_y}]) + 
(-1)^{L_y}[2,3,...,L_y|h_2,h_3,...,h_{L_y}] \cr\cr
& = &
(-1)^{L_y-1}((q-2)[1,2,...,L_y-1|h_1,h_2,...,h_{L_y-1}] + ...  +
(q-2)[2,3,...,L_y|h_2,h_3,...,h_{L_y}] \ . \cr\cr
& &
\label{lastrowtri}
\eeqs
The theorem then follows.  $\Box$

\medskip

Thus, for example, for $L_y=5$, 
\beq
\bar T^{(5,4)}=\left(
  \begin{array}{ccccc}
   q-3 & -1  &  0  &  0  &  0  \cr
   q-4 & q-4 & -1  &  0  &  0  \cr
   q-4 & q-4 & q-4 & -1  &  0  \cr
   q-4 & q-4 & q-4 & q-4 & -1  \cr
   q-2 & q-2 & q-2 & q-2 & q-2 
    \end{array}\right ) \ .
\label{tbartri54}
\eeq
We also recall the result from \cite{pm,tor4} that the $\lambda_{P,G,j}$'s for
the cyclic and M\"obius strips graphs $G$ of a given lattice and width $L_y$
are the same.  

Here the eigenvalues of the cyclic triangular lattice do not have a simple
linear form $q-a_{L_y,j}$ as was the case for the eigenvalues for this degree
$d=L_y-1$ in the case of the cyclic and M\"obius strips of the square lattice
given in eqs. (\ref{lamdlym1}) and (\ref{asqlyj}) or the eigenvalues for the
self-dual strip of the square lattice given in eq. (\ref{lamdbc2dlym1}) with
(\ref{adbc2lyj}).

However, we do find the following corollary, which is easily derived from
the theorem. 

\medskip

{\bf Corollary 7} \quad  For the cyclic strip of the triangular
lattice 
\beq
{\rm det}(\bar T^{(L_y,L_y-1)}) = (q-2)^2(q-3)^{L_y-2}
\label{lamprodtri}
\eeq
\beq
{\rm Tr}(\bar T^{(L_y,L_y-1)}) = 3+L_y(q-4) \ .
\label{lamsumtri}
\eeq

For example, for $L_y=2$, we have, for the eigenvalue of the reduced transfer
matrix, 
\beq
\lambda_{2,1,j} = \frac{1}{2}\biggl [ 2q-5 \pm \sqrt{9-4q} \ \biggr ] \quad
{\rm for} \ j=1,2
\label{lamtri21j}
\eeq
in agreement with eqs. (5.12) and (5.13) of \cite{wcy} (taking account of
eq. (\ref{tbardef})).  For $L_y=3$, we have
\beq 
\lambda_{3,2,1} = q-2
\label{lam32jtri}
\eeq
\beq
\lambda_{3,2,j} = \frac{1}{2}\biggl [ 2q-7 \pm \sqrt{25-8q} \ \biggr ]
\quad {\rm for} \ j=2,3 
\label{lamtri32j} 
\eeq
in agreement our eqs. (2.16) and (2.17) in \cite{t}.

\section{Chromatic Polynomial for the Cyclic Strip of the Square Lattice of
Width $L_y=5$}

\subsection{General} 

In this section we give our solution for the chromatic polynomial of the $L_y
\times L_x$ cyclic strip of the square lattice with width $L_y=5$.  For $L_x$
beyond the first few degenerate cases, this has $n=L_xL_y$ vertices and
$e=L_x(2L_y-1)$ edges.  The chromatic number is $\chi=2$ for $L_x$ even and
$\chi=3$ for $L_x$ odd, independent of $L_y$.

We list below the specific $c^{(d)}$'s that will appear in our results. 
\beq
c^{(0)}=1 \ , \quad c^{(1)}=q-1 \ , \quad c^{(2)}=q^2-3q+1 \ ,
\label{cd012}
\eeq
\beq
c^{(3)}=q^3-5q^2+6q-1 \ ,
\label{cd3}
\eeq
\beq
c^{(4)}=(q-1)(q^3-6q^2+9q-1) \ , 
\label{cd4}
\eeq
and
\beq
c^{(5)}=q^5-9q^4+28q^3-35q^2+15q-1 \ . 
\label{cd5}
\eeq

For this family, using the general structural formulas derived in \cite{cf}, we
know that
\beq
n_P(5,0)=M_4=9 \ , \quad n_P(5,1)=n_P(5,2)=21 \ , \quad n_P(5,3)=13 \ , \quad
n_P(5,4)=5
\label{np5d}
\eeq
as well as $n_P(5,5)=1$.  Summing these or equivalently evaluating the general
formula (\ref{nptot}) for the case $L_y=5$, one has, for the total number of
$\lambda_{5,d,j}$'s, $N_{P,5,\lambda}=70$.  We have calculated the chromatic
polynomials by first computing the transfer matrices $T_{L_y,d}$ for $0 \le d
\le L_y$.

We find 
\beqs
P(sq[5 \times m,cyc.],q) & = & 
c^{(0)} \sum_{j=1}^9 (\lambda_{5,0,j})^m + 
c^{(1)} \sum_{j=1}^{21} (\lambda_{5,1,j})^m + 
c^{(2)} \sum_{j=1}^{21} (\lambda_{5,2,j})^m \cr\cr & + &
c^{(3)} \sum_{j=1}^{13} (\lambda_{5,3,j})^m + 
c^{(4)} \sum_{j=1}^5 (\lambda_{5,4,j})^m + c^{(5)}(\lambda_{5,5})^m
\label{ply5cyc}
\eeqs
where 
$\lambda_{5,0,j}$ for $j=1,2$ and $3 \le j \le 9$ are roots of
equations of degree 2 and 7 in $q$, 
$\lambda_{5,1,j}$ for $1 \le j \le 8$ and $9 \le j \le 21$ are roots 
of equations of respective degrees 8 and 13, 
$\lambda_{5,2,j}$ for $1 \le j \le 9$ and $10 \le j \le 21$ are 
roots of equations of respective degrees 9 and 12, and 
$\lambda_{5,3,j}$ for $1 \le j \le 5$ and $6 \le j \le 13$ are roots of 
equations of respective degrees 5 and 8.  We discuss these $\lambda$'s next. 

\subsection{$\lambda_{5,0,j}$}

The quadratic equations for $\lambda_{5,0,j}$ factorize over the field
${\mathbb Q}[\sqrt{5}]$, so that these eigenvalues are polynomials in $q$ and
the elements of this field:
\beq
\lambda_{5,0,j} = \frac{1}{2}\bigg [ 2q^3-13q^2+28q-19 \pm (q-1)(q-3)\sqrt{5}
\bigg ] \quad {\rm for} \quad j=1,2 \ . 
\label{lam5012}
\eeq
The $\lambda_{5,0,j}$, $3 \le j \le 9$ are identical to the $\lambda$'s for the
$L_y=5$ strip with free boundary conditions, and the degree-7 equation for
these is 
\beq
\xi^7 + \sum_{j=1}^7 b_{sq5FF,j}\xi^{7-j} = 0
\label{bsq5ff}
\eeq
where the expressions $ b_{sq5FF,j}$ were given in eqs. (A.20)-(A.26) of
\cite{s4}. 

\subsection{$\lambda_{5,d,j}$, $d=1,2$}

The $\lambda_{5,d,j}$ with $d=1$ and $d=2$ and $1 \le j \le 21$ are too
complicated to present here.  The computer files are available on request from
the authors. 

\subsection{$\lambda_{5,3,j}$}

The $\lambda_{5,3,j}$ are the eigenvalues of the transfer matrix 
\beq
T^{(5,3)} = \left( \begin{array}{ccccccccccccc}
     t_{11} & q-2 & -1 & 0 & 0 & 0 & 0 & 0 & 0 & 0 & 1 & 0 & 0 \\
     q-2 & t_{22} & q-3 & q-2 & -1 & 0 & 0 & 0 & 0 & 0 & 1 & 0 & 0 \\
     -1 & q-3 & t_{33} & -1 & q-3 & 0 & 0 & -1 & 0 & 0 & 1 & 1 & 0 \\
     0 & q-2 & -1 & t_{44} & q-3 & q-2 & -1 & 0 & 0 & 0 & 0 & 0 & 0 \\
     0 & -1 & q-3 & q-3 & t_{55} & -1 & q-3 & q-3 & -1 & 0 & 0 & 1 & 0 \\
     0 & 0 & 0 & q-2 & -1 & t_{66} & q-2 & 0 & 0 & 0 & 0 & 0 & 0 \\
     0 & 0 & 0 & -1 & q-3 & q-2 & t_{77} & -1 & q-2 & 0 & 0 & 0 & 0 \\
     0 & 0 & -1 & 0 & q-3 & 0 & -1 & t_{88} & q-3 & -1 & 0 & 1 & 1 \\
     0 & 0 & 0 & 0 & -1 & 0 & q-2 & q-3 & t_{99} & q-2 & 0 & 0 & 1 \\
     0 & 0 & 0 & 0 & 0 & 0 & 0 & -1 & q-2 & t_{1010} & 0 & 0 & 1 \\
     2-q & 1 & 2-q & 0 & 0 & 0 & 0 & 0 & 0 & 0 & q-2 & 0 & 0 \\
     0 & 0 & 2-q & 0 & 1 & 0 & 0 & 2-q & 0 & 0 & 0 & q-2 & 0 \\
     0 & 0 & 0 & 0 & 0 & 0 & 0 & 2-q & 1 & 2-q & 0 & 0 & q-2
\end{array} \right) \\
\eeq
where
\beqs
t_{11} = -q^2+4q-5, \quad & t_{22} = -(q-2)(q-3), \quad & t_{33} =
-q^2+5q-8 \\ \nonumber
t_{44} = -(q-2)(q-3), \quad & t_{55} = -(q-3)^2, \quad & t_{66} =
-(q-2)^2 \\ \nonumber
t_{77} = -(q-2)(q-3), \quad & t_{88} = -q^2+5q-8, \quad & t_{99} =
-(q-2)(q-3) \\ \nonumber
t_{1010} = -q^2+4q-5 \ . & & 
\eeqs
As noted above, the characteristic polynomial of this matrix factorizes into
degree-8 and degree-5 polynomials. 

\subsection{$\lambda_{5,4,j}$ and $\lambda_{5,5}$}

As a special case of our general formula (\ref{lamdlym1}) with (\ref{asqlyj})
for $\lambda_{L_y,L_y-1,j}$, of which there are $L_y$ eigenvalues, we have
\beq
\lambda_{5,4,1}=q-a_{sq,5,1}=q-1
\label{lam541}
\eeq
\beq
\lambda_{5,4,2}=q-a_{sq,5,2}=q-\left ( \frac{5-\sqrt{5} \ }{2} \right ) = 
q-1.38196...
\label{lam542}
\eeq
\beq
\lambda_{5,4,3}=q-a_{sq,5,3}=q-\left ( \frac{7-\sqrt{5} \ }{2} \right ) = 
q-2.38196...
\label{lam543}
\eeq
\beq
\lambda_{5,4,4}=q-a_{sq,5,4}=q-\left ( \frac{5+\sqrt{5} \ }{2} \right ) = 
q-3.61803...
\label{lam544}
\eeq
\beq
\lambda_{5,4,5}=q-a_{sq,5,5}=q-\left ( \frac{7+\sqrt{5} \ }{2} \right ) = 
q-4.61803...
\label{lam545}
\eeq
We note the connection between the appearance of the algebraic numbers 
$a+b\sqrt{5}$, i.e., elements of the field ${\mathbb Q}[\sqrt{5}]$ in both the
$\lambda_{5,4,j}$ for $2 \le j \le 5$ and in $\lambda_{5,0,\ell}$ for
$\ell=1,2$.  Indeed, the term involving $\sqrt{5}$ in $\lambda_{5,0,1}$ and 
$\lambda_{5,0,2}$ is proportional to $(q-1)(q-3)$, which is $\lambda_{4,0,1}$. 
Our general formula (\ref{lamlyly}) yields 
\beq
\lambda_{5,5}=-1 \ . 
\label{lam55}
\eeq

\section{Chromatic Polynomial for the M\"obius Strip of the Square Lattice of
Width $L_y=5$}

We obtain this by applying the transformation rules derived in \cite{cf} to go
from the chromatic polynomial for the strip with width $L_y$ and arbitrary 
length $L_x$ with $(FBC_y,PBC_x)=$ cyclic boundary conditions to the
corresponding strip with $(FBC_y,TPBC_x)=$ M\"obius boundary conditions. 
For the chromatic polynomial of the M\"obius strip with width $L_y=5$, we find
\beqs
& & P(sq[5 \times m,Mb],q) = c^{(0)} \bigg [ \sum_{j=1}^7 
(\lambda_{5,0,j+2})^m +
           \sum_{j=1}^9 (\lambda_{5,2,j})^m - 
           \sum_{j=1}^2 (\lambda_{5,0,j})^m - 
           \sum_{j=1}^{12} (\lambda_{5,2,j+9})^m \bigg ] \cr\cr
& + & c^{(1)}\bigg [- \sum_{j=1}^8 (\lambda_{5,1,j+5})^m + 
                    \sum_{j=1}^{13} (\lambda_{5,1,j+8})^m +
                    \sum_{j=1}^5 (-1)^j (\lambda_{5,4,j})^m \bigg ] \cr\cr
& + & c^{(2)}\bigg [ \sum_{j=1}^8 (\lambda_{5,3,j+5})^m - 
 \sum_{j=1}^5 (\lambda_{5,3,j})^m \bigg ] + c^{(3)}(\lambda_{5,5})^m \ . 
\label{ply5mb}
\eeqs
Expressing this in the general form (\ref{pgsummb}), we have 
\beqs
P(sq[5 \times m,Mb],q) & = &
c^{(0)}\bigg [ \sum_{j=1}^{16} (\lambda_{5,0,+,j})^m
-\sum_{j=1}^{14}(\lambda_{5,0,-,j})^m \bigg ] 
+ c^{(1)} \bigg [ \sum_{j=1}^{15} (\lambda_{5,1,+,j})^m - 
\sum_{j=1}^{11}(\lambda_{5,1,-,j})^m \bigg ] \cr\cr
& + & c^{(2)}\bigg [ \sum_{j=1}^8 (\lambda_{5,2,+,j})^m - 
\sum_{j=1}^5 (\lambda_{5,2,-,j})^m \bigg ]
+ c^{(3)}(\lambda_{5,5})^m
\label{ply5mbgen}
\eeqs
where
\beq
\lambda_{5,0,+,j} = \lambda_{5,0,j+2} \quad {\rm for} \quad 1 \le j \le 7
\label{lam50plusj1to7}
\eeq
\beq
\lambda_{5,0,+,j} = \lambda_{5,2,j-7} \quad {\rm for} \quad 8 \le j \le 16
\label{lam50plusj8to16}
\eeq
\beq
\lambda_{5,0,-,j} = \lambda_{5,0,j} \quad {\rm for} \quad 1 \le j \le 2
\label{lam50minus1to2}
\eeq
\beq
\lambda_{5,0,-,j} = \lambda_{5,2,j+7} \quad {\rm for} \quad 3 \le j \le 14
\label{lam50minus3to14}
\eeq
\beq
\lambda_{5,1,+,j} = \lambda_{5,1,j+8} \quad {\rm for} \quad 1 \le j \le 13
\label{lam51plusj1to13}
\eeq
\beq
\lambda_{5,1,+,14} = \lambda_{5,4,2} \ , \quad 
\lambda_{5,1,+,15} = \lambda_{5,4,4} 
\label{lam51plusj14to15}
\eeq
\beq
\lambda_{5,1,-,j} = \lambda_{5,1,j} \quad {\rm for} \quad 1 \le j \le 8
\label{lam51minus1to8}
\eeq
\beq
\lambda_{5,1,-,9} =  \lambda_{5,4,1}, \quad
\lambda_{5,1,-,10} = \lambda_{5,4,3}, \quad
\lambda_{5,1,-,11} = \lambda_{5,4,5}
\label{lam51minus9to11}
\eeq
\beq
\lambda_{5,2,+,j} = \lambda_{5,3,j+5} \quad {\rm for} \quad 1 \le j \le 8
\label{lam52plusj1to8}
\eeq
\beq
\lambda_{5,2,-,j} = \lambda_{5,3,j} \quad {\rm for} \quad 1 \le j \le 5
\label{lam52minusj1to8}
\eeq
\beq
\lambda_{5,3,+,j} = \lambda_{5,5} \ . 
\label{lam53plusj1}
\eeq

\section{Locus ${\cal B}$ for $L_y=5$ Cyclic/M\"obius Strips of the Square 
Lattice } 

In this section we present the singular boundary ${\cal B}$ across which $W(q)$
is singular in the complex $q$ plane for the $L_x \to \infty$ limit of the
cyclic strip of the square lattice with width $L_y=5$.  We also recall that
that ${\cal B}$ is the same for cyclic and M\"obius strips
\cite{w,wcyl,wcy,pm,s4}, as was proved in \cite{tor4}.  A plot of ${\cal B}$ is
shown in Fig. \ref{sqpxy5}

\begin{figure}[hbtp]
\centering
\leavevmode
\epsfxsize=4in
\begin{center}
\leavevmode
\epsffile{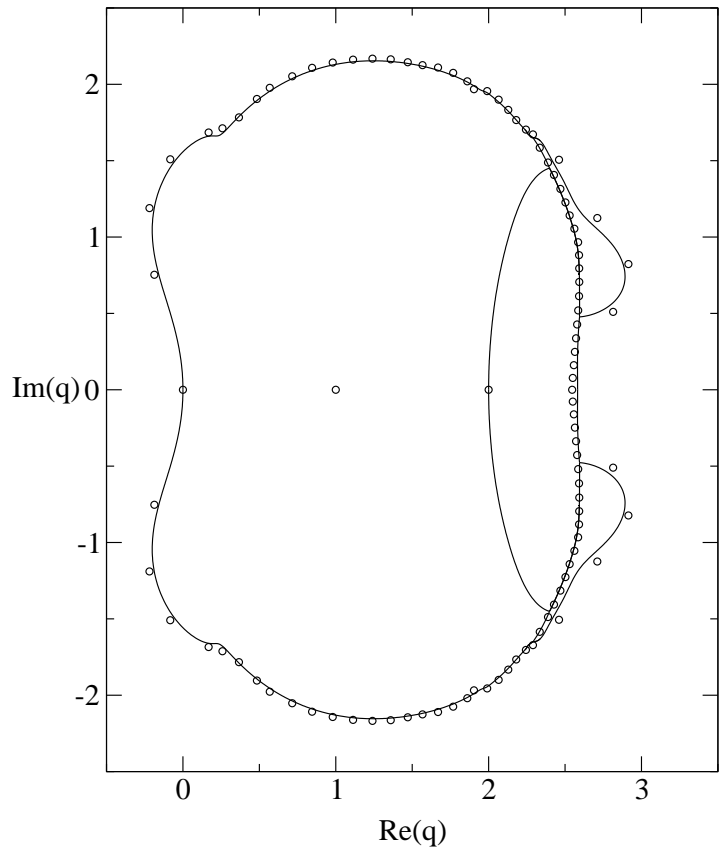}
\end{center}
\caption{\footnotesize{Locus ${\cal B}$ for the $L_x \to \infty$ limit of the
strip of the square lattice of width $L_y=5$ strip with $(FBC_y,(T)PBC_x)=$
cyclic (equiv. M\"obius) boundary conditions.  For comparison, chromatic zeros
calculated for the strip length $L_x=20$ (i.e., $n=100$ vertices) are shown.}}
\label{sqpxy5}
\end{figure}

The locus ${\cal B}$ is comprised of closed curves that separate the $q$
plane into several regions.  This locus crosses the real axis at $q=0$, $q=2$,
and a maximal value, $q_c$, which is
\beq
q_c = 2.582385... 
\label{qcsqcyc}
\eeq
The region $R_1$ includes the real axis for $q \ge q_c$ and $q \le 0$ and
extends outward to complex infinity from the outer envelope of ${\cal B}$.  The
region $R_2$ includes the real segment $2 \le q \le q_c$, while region $R_3$
includes the real segment $0 \le q \le 2$.  In region $R_1$, the dominant
$\lambda$ is the root of the degree-7 equation obtained from the appendix of
\cite{s4} with the largest magnitude, which we label $\lambda_{R1}$.  In region
$R_2$, the dominant $\lambda_j$ is the root of the degree-12 equation with the
largest magnitude, which we label $\lambda_{R2}$.  Hence, $q_c$ is the solution
of the equation of the degeneracy $|\lambda_{R1}|=|\lambda_{R2}|$.  The value
of $q_c$ that we have obtained for the present strip may be compared with the
values that we obtained previously for the cyclic (M\"obius) $L_y \times
\infty$ strips of the square lattice with smaller widths, namely, $q_c=2$ for
$L_y=1,2$ \cite{w}, $q_c = 2.33654...$ for $L_y=3$ \cite{wcyl,wcy}, and $q_c =
2.492845...$ for $L_y=4$ \cite{s4}.  We calculate that $W(sq[5 \times \infty,
FBC_y,(T)PBC_x],q)=1.317594$ at $q_c$, as given in eq. (\ref{qcsqcyc}).  In
region $R_3$, the dominant $\lambda_j$ is the root of the degree-13 equation
with the largest magnitude, which we label $\lambda_{R3}$.  There are also
complex-conjugate regions centered at approximately $q=2.7 \pm 0.8i$; we denote
these as $R_4, R_4^*$.  The dominant $\lambda_j$ in these regions is the root
of the degree-13 equation of maximal magnitude here; this is denoted
$\lambda_{R4}$. We have
\beq
W=(\lambda_{R1})^{1/5} \ , \quad {\rm for} \quad q \in R_1
\label{wr1}
\eeq
\beq
|W|=|\lambda_{Rj}|^{1/5} \ , \quad {\rm for} \quad q \in R_j \ , \ \ j=2,3,4 
\ .
\label{wrother}
\eeq
(In regions other than $R_1$, only the magnitude $|W|$ can be determined
unambiguously \cite{w}.)  Our previous results such as \cite{wcyl,wcy} showed
that there can also be very small sliver regions in the complex $q$ plane; we
have not made an exhaustive search for these.  As we found previously for
$L_y=3$ \cite{wcyl,wcy} and $L_y=4$ \cite{s4}, the locus ${\cal B}$ has support
for $Re(q) < 0$ as well as $Re(q) \ge 0$.

Fig. \ref{sqpxy5} also shows a comparison of the chromatic zeros calculated for
a long finite cyclic strip with $L_x=20$, i.e., $n=100$ vertices, versus the
asymptotic locus ${\cal B}$.  One sees that, aside from the isolated real zero
at $q=1$, these chromatic zeros lie reasonably close to the locus ${\cal B}$.
On the curve forming ${\cal B}$ passing through $q=0$ and $q=q_c$, the density
is highest on the right-hand side and somewhat lower in the region of $q=0$.
The density of chromatic zeros on the part of ${\cal B}$ passing through $q=2$
and ending at upper and lower triple points is very low, while one observes an
intermediate density on the complex-conjugate pair of bulb-like curves
protruding to the right.  These features are qualitatively the same as we found
for $L_y=3$ (Fig. 1 of \cite{wcyl}) and for $L_y=4$ (Fig. 1 of \cite{s4}).  The
differences in densities on different portions of ${\cal B}$ were less
pronounced for $L_y=2$ (Fig. 1 of \cite{w}).  The elementary case $L_y=1$ is 
special in two respects: (i) the complex chromatic zeros lie exactly ${\cal B}$
for finite as well as infinite $L_x$, and (ii) their density is constant on
${\cal B}$ \cite{w}.

\section{Discussion}

Here we make several further comments on these results for ${\cal B}$ 

\begin{enumerate}

\item 

As in our earlier work, we characterize a strip graph as containing a global
circuit if it contains a cycle along the longitudinal direction whose length
goes to infinity as the length $L_x \to \infty$; in practice, this is
equivalent to the property that the lattice strip graph has periodic
longitudinal boundary conditions.  For all of the strips of the square lattice
containing global circuits that we have studied, the locus ${\cal B}$ encloses
regions of the $q$ plane including certain intervals on the real axis and
passes through $q=0$ and $q=2$ as well as other possible points, depending on
the family.  Note that the presence of global circuits is a sufficient, but not
necessary, condition for ${\cal B}$ to enclose regions, as was shown in
\cite{strip2} (see Fig. 4 of that work).  Our present results for the square
lattice are in accord with, and strengthen the evidence for, the inference
(conjecture) \cite{bcc,a} that
\beq
\quad {\cal B} \supset \{q=0, \ 2 \} \ \ {\rm for}
\ \ sq[L_y,FBC_y,(T)PBC_x] \quad \forall \ L_y \ge 1 \ . 
\label{bcrossq02}
\eeq 
(Of course $L_y=1$ graphs with $(FBC_y,TPBC_x)$ and $(FBC_y,PBC_x)$
boundary conditions are identical.)

\item 

The crossing of ${\cal B}$ at the point $q=2$ for the (infinite-length limit
of) strips with global circuits nicely signals the existence of a
zero-temperature critical point in the Ising antiferromagnet (equivalent to the
Ising ferromagnet on bipartite graphs). This has been discussed in \cite{a} in
the context of exact solutions for finite-temperature Potts model partition
functions on the $L_y=2$ cyclic and M\"obius strips of the square lattice.  In
contrast, this connection is not, in general, present for strips with free
longitudinal boundary conditions since ${\cal B}$ does not, in general, pass
through $q=2$. Furthermore, for the strips without
global circuits, there is no indication of any motion of the respective loci
${\cal B}$ toward $q=2$ as $L_y$ increases. 

\item 

For cyclic strips, we note a correlation between the coefficient $c_{G_s,j}$ of
the respective dominant $\lambda_{G_s,j}$'s in regions that include intervals
of the real axis.  Before, it was shown \cite{bcc} that the $c_{G_s,j}$ of the
dominant $\lambda_{G_s,j}$ in region $R_1$ including the real intervals $q >
q_c(\{G\})$ and $q < 0$ is $c^{(0)}=1$, where the $c^{(d)}$ were given in
eqs. (\ref{cd}).  We observe further that the $c_{G_s,j}$ that multiplies the
dominant $\lambda_{G_s,j}$ in the region containing the intervals $0 < q < 2$
is $c^{(1)}$.  For the cyclic $L_y=3$ and $L_y=4$ strips, there is also another
region containing an interval $2 \le q \le q_c$ on the real axis, where $q_c
\simeq 2.34$ and 2.49 for $L_y=3,4$; in this region, we find that the
$c_{G_s,j}$ multiplying the dominant $\lambda_{G_s,j}$ is $c^{(2)}$.  Our
present results again agree with this conjecture.

\item

For the $L_x \to \infty$ limit of all of the strips of the square lattice
containing global circuits, a $q_c$ is defined, and our results for the cyclic
and M\"obius strips with widths from $L_y=1$ through $L_y=4$ indicate that
$q_c$ is a non-decreasing function of $L_y$ in these cases.  The same behavior
was found for the strips of the triangular lattice with $L_y=2$ \cite{wcy} (and
subsequently also $L_y=3,4$ \cite{t}). 
This motivated the inference (conjecture) that $q_c$ is a
non-decreasing function of $L_y$ for strips of regular lattices with
$(FBC_y,(T)PBC_x)$ boundary conditions \cite{bcc}, and our present results
strengthen the support for this inference.  Given that, as $L_y \to \infty$,
$q_c$ reaches a limit, which is the $q_c$ for the 2D lattice of the specified
type (square, triangular, etc.), this inference leads to the following 
inequality: 
\beq 
q_c(\Lambda, L_y \times \infty,BC_y,(T)PBC_x) \le q_c(\Lambda) \ . 
\label{qcineq}
\eeq
This is a non-decreasing function of $L_y$.  Our current results are in 
agreement with this conjecture. 

\item 

One interesting feature of our new results is that the outermost envelope of
the curves on ${\cal B}$ for the $L_y=5$ strip of the square lattice do not lie
outside the envelope for ${\cal B}$ for the $L_y=4$ strip.  This is evident
from the comparison shown in Fig. \ref{sqpxy45}.  For reference, we also show
the analogous pairwise comparisons for widths $L_y=2,3$ and $L_y=3,4$.  For
each of these, it was true that as $L_y$ increases, the outer envelope of the
locus ${\cal B}$ moves outward in a manner such that ${\cal B}$ for width
$L_y+1$ encloses the locus ${\cal B}$ for $L_y$, allowing for equality at some
points, in particular, $q=0$ \cite{bcc}.  This is also true for the pairwise
comparison of the loci ${\cal B}$ for $L_y=1$ and $L_y=2$, as is clear from
Fig. \ref{sqpxy23} and the fact that the locus for $L_y=1$ is the circle
$|q-1|=1$.  However, our new results show that this behavior does not persist
in the comparison of $L_y=4$ and $L_y=5$ and hence is not general.

\end{enumerate} 

\begin{figure}[hbtp]
\centering
\leavevmode
\epsfxsize=4in
\begin{center}
\leavevmode
\epsffile{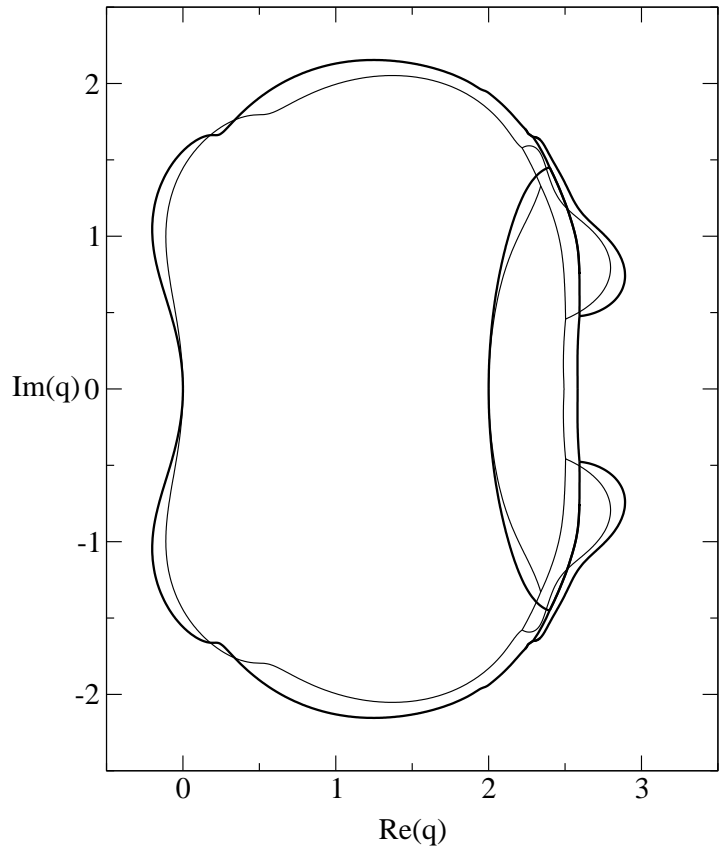}
\end{center}
\caption{\footnotesize{Comparison of loci ${\cal B}$ for the $L_x \to \infty$ 
limits of the strips of the square lattice of widths $L_y=4$ (light curves) and
$L_y=5$ (heavy curves) with $(FBC_y,(T)PBC_x)=$ cyclic (equiv. M\"obius) 
boundary conditions.}}
\label{sqpxy45}
\end{figure}

\begin{figure}[hbtp]
\centering
\leavevmode
\epsfxsize=4in
\begin{center}
\leavevmode
\epsffile{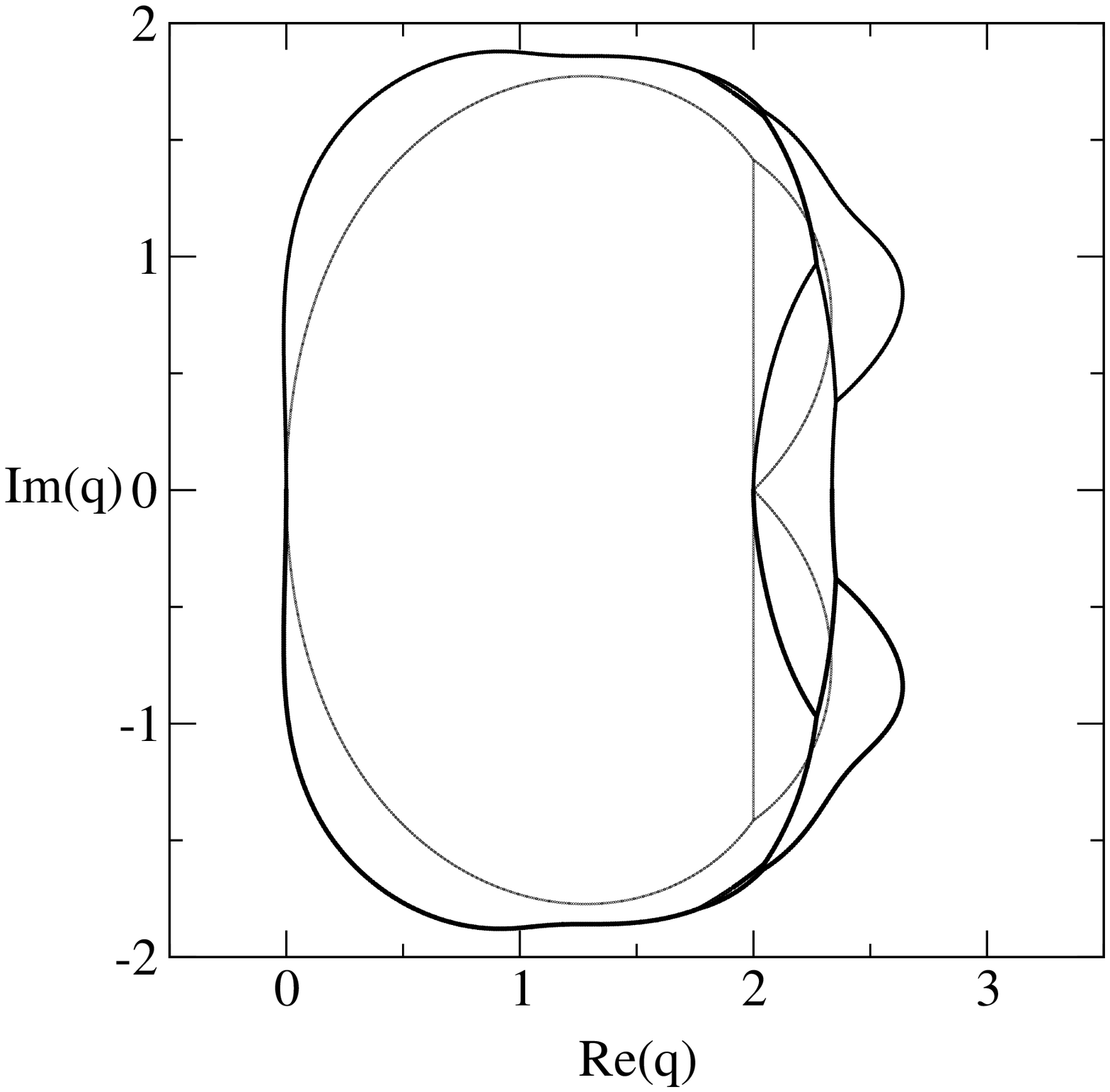}
\end{center}
\caption{\footnotesize{Comparison of loci ${\cal B}$ for the $L_x \to \infty$
limits of the strips of the square lattice of widths $L_y=2$ (light curves) and
$L_y=3$ (heavy curves) with $(FBC_y,(T)PBC_x)=$ cyclic (equiv. M\"obius)
boundary conditions.}}
\label{sqpxy23}
\end{figure}

\begin{figure}[hbtp]
\centering
\leavevmode
\epsfxsize=4in
\begin{center}
\leavevmode
\epsffile{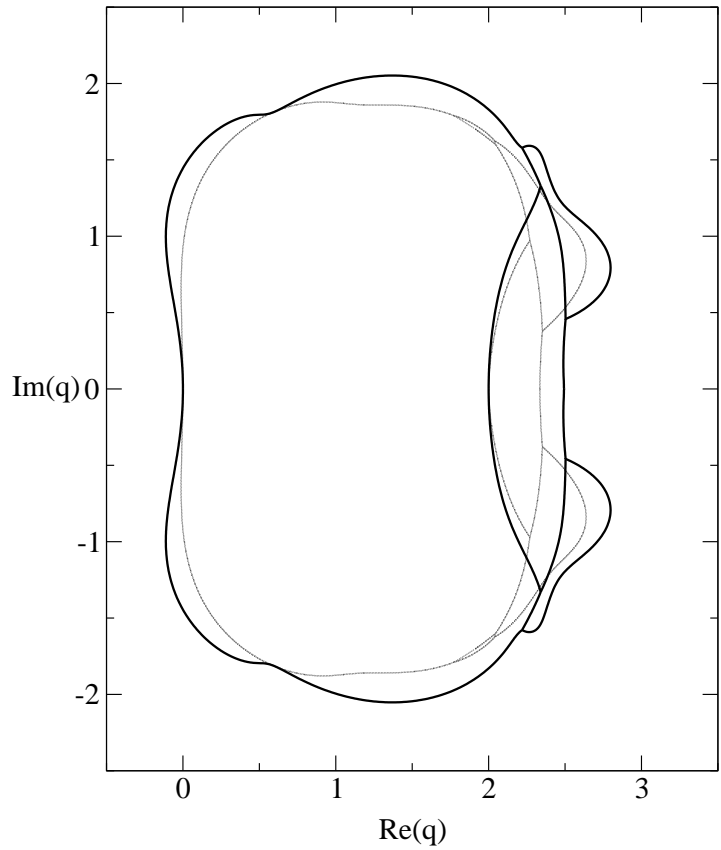}
\end{center}
\caption{\footnotesize{Comparison of loci ${\cal B}$ for the $L_x \to \infty$
limits of the strips of the square lattice of widths $L_y=3$ (light curves) and
$L_y=4$ (heavy curves) with $(FBC_y,(T)PBC_x)=$ cyclic (equiv. M\"obius)
boundary conditions.}}
\label{sqpxy34}
\end{figure}

\vspace{10mm}

Acknowledgment: The research of R. S. was supported in part by the NSF 
grant PHY-0098527.

\section{Appendix: Structure of Chromatic Chromatic Polynomials for Cyclic 
and M\"obius Strips of the Square Lattice}

\subsection{$L_y=2$ Cyclic Strip}

We list some known results to illustrate our general notation, for comparison
with our new calculation of the $L_y=5$ case.  The chromatic polynomial of the
cyclic strip with width $L_y=2$ is \cite{bds}
\beqs
P(2 \times m,cyc.,q) & = & 
c^{(0)}(\lambda_{2,0,1})^m + c^{(1)}\sum_{j=1}^2 (\lambda_{2,1,j})^m 
+ c^{(2)}(\lambda_{2,2})^m \cr\cr
& = & (q^2-3q+3)^m + (q-1)\Big [ (1-q)^m + (3-q)^m \Big ] +
(q^2-3q+1) \cr\cr
& &
\label{ply2cyc}
\eeqs
where
\beq
\lambda_{2,0,1}=q^2-3q+3
\label{lam201}
\eeq
\beq
\lambda_{2,1,1}=-(q-a_{2,1})=1-q
\label{lam211}
\eeq
\beq
\lambda_{2,1,2}=-(q-a_{2,2})=3-q
\label{lam212}
\eeq
\beq
\lambda_{2,2}=1 \ .
\label{lam22}
\eeq

\subsection{$L_y=2$ M\"obius Strip}

The chromatic polynomial of the M\"obius ($Mb$) strip with width $L_y=2$ is
\cite{bds}
\beqs
P(2 \times m,Mb,q) & = & 
c^{(0)} \bigg [ (\lambda_{2,0,1})^m - (\lambda_{2,2})^m \bigg ] +
c^{(1)} \sum_{j=1}^2 (-1)^j (\lambda_{2,1,j})^m \cr\cr
& = & \Big [ (q^2-3q+3)^m -1 \Big ] + (q-1)\Big [-(1-q)^m + (3-q)^m \Big ]
\ .
\label{ply2mb}
\eeqs
Expressing this in the general form (\ref{pgsummb}), we have
\beq
P(2 \times m,Mb,q) = 
c^{(0)}\bigg [ (\lambda_{2,0,+,1})^m - (\lambda_{2,0,-,1})^m \bigg ] + 
c^{(1)}\bigg [ (\lambda_{2,1,+,1})^m - (\lambda_{2,1,-,1})^m \bigg ]
\label{ply2mbgen}
\eeq
where
\beq
\lambda_{2,0,+,1}=\lambda_{2,0,1}
\label{lam20plus1}
\eeq
\beq
\lambda_{2,0,-,1}=\lambda_{2,2}
\label{lam20minus1}
\eeq
\beq
\lambda_{2,1,+,1}=\lambda_{2,1,2}
\label{lam21plus1}
\eeq
\beq
\lambda_{2,1,-,1}=\lambda_{2,1,1} \ .
\label{lam21minus1}
\eeq

\subsection{$L_y=3$ Cyclic Strip}

The chromatic polynomial for the cyclic strip of width $L_y=3$ is
\cite{wcyl,wcy} 
\beqs
P(3 \times m,cyc.,q) & = & 
c^{(0)}\sum_{j=1}^2 (\lambda_{3,0,j})^m +
c^{(1)}\sum_{j=1}^4 (\lambda_{3,1,j})^m + 
c^{(2)}\sum_{j=1}^3 (\lambda_{3,2,j})^m \cr\cr & + & 
c^{(3)}(\lambda_{3,3})^m
\label{ply3cyc}
\eeqs
where $\lambda_{3,0,j}$ for $j=1,2$ are the solutions to a quadratic equation
given in \cite{wcy}, 
\beq
\lambda_{3,1,1}=-(q-2)^2
\label{lam311}
\eeq
the $\lambda_{3,1,j}$ for $j=4,5,6$ are roots of a cubic equation given
in \cite{wcy}, and $\lambda_{3,2,j}=q-a_{3,j}$, with the specific values 
\beq
\lambda_{3,2,1}=q-a_{3,1}=q-1
\label{lam321}
\eeq
\beq
\lambda_{3,2,2}=q-a_{3,2}=q-2
\label{lam322}
\eeq
\beq
\lambda_{3,2,3}=q-a_{3,3}=q-4 \ .
\label{lam323}
\eeq
Finally, 
\beq
\lambda_{3,3}=-1 \ .
\label{lam33}
\eeq

\subsection{$L_y=3$ M\"obius Strip}

The chromatic polynomial for the M\"obius strip of width $L_y=3$ is \cite{pm}
\beqs
P(3 \times m,Mb,q) & = & 
c^{(0)} \bigg [ \sum_{j=1}^2 (\lambda_{3,0,j})^m + 
\sum_{j=1}^3 (-1)^j (\lambda_{3,2,j})^m \bigg ] \cr\cr
& + & c^{(1)} \bigg [\sum_{j=1}^3 (\lambda_{3,1,j+1})^m-(\lambda_{3,1,1})^m 
\bigg ] + c^{(2)}(\lambda_{3,3})^m \ .
\label{ply3mb}
\eeqs
Expressing this in the general form (\ref{pgsummb}), 
\beqs
P(3 \times m,Mb,q) & = &
c^{(0)}\bigg [ \sum_{j=1}^3 (\lambda_{3,0,+,j})^m 
- \sum_{j=1}^2(\lambda_{3,0,+,j})^m \bigg ]
\cr\cr 
& + & c^{(1)} \bigg [ \sum_{j=1}^3 (\lambda_{3,1,+,j})^m 
- (\lambda_{3,1,-,1})^m \bigg ] + c^{(2)}(\lambda_{3,3})^m
\label{ply3mbgen}
\eeqs
where
\beq
\lambda_{3,0,+,j} = \lambda_{3,0,j} \quad {\rm for} \quad j=1,2
\label{lam30plusj}
\eeq
\beq
\lambda_{3,0,+,3}=\lambda_{3,2,2}
\label{lam30plus3}
\eeq
\beq
\lambda_{3,0,-,1}=\lambda_{3,2,1}
\label{lam30minus1}
\eeq
\beq
\lambda_{3,0,-,2}=\lambda_{3,2,3}
\label{lam30minus2}
\eeq
\beq
\lambda_{3,1,+,j}=\lambda_{3,1,j+1} \quad {\rm for} \quad j=1,2,3
\label{lam31plus13}
\eeq
\beq
\lambda_{3,1,-,1}=\lambda_{3,1,1}
\label{lam31minus1}
\eeq
\beq
\lambda_{3,2,+,1} =\lambda_{3,3} \ .
\label{lam32plus1}
\eeq

\subsection{$L_y=4$ Cyclic Strip}

The chromatic polynomial of the cyclic strip of width $L_y=4$ is \cite{s4}
\beqs
P(4 \times m,cyc.) & = & 
c^{(0)} \sum_{j=1}^4 (\lambda_{4,0,j})^m + 
c^{(1)}\sum_{j=1}^9 (\lambda_{4,1,j})^m + 
c^{(2)} \sum_{j=1}^8 (\lambda_{4,2,j})^m \cr\cr & + & 
c^{(3)} \sum_{j=1}^4 (\lambda_{4,3,j})^m + 
c^{(4)}(\lambda_{4,4})^m
\label{ply4cyc}
\eeqs
where
\beq
\lambda_{4,0,1}=(q-1)(q-3)
\label{lam401}
\eeq
$\lambda_{4,0,j}$ for $j=2,3,4$ are roots of a cubic equation, 
$\lambda_{4,1,j}$ for $1 \le j \le 4$ and $5 \le j \le 9$ are roots of
equations of respective degrees 4 and 5, 
$\lambda_{4,2,j}$ for $1 \le j \le 3$ and $4 \le j \le 8$ are roots of
equations of respective degrees 3 and 5, all given in \cite{s4}, and 
$\lambda_{4,3,j}=a_{4,j}-q$, with specific values 
\beq
\lambda_{4,3,1}=-(q-a_{4,1})=1-q
\label{lam431}
\eeq
\beq
\lambda_{4,3,2}=-(q-a_{4,2})=3-\sqrt{2}-q
\label{lam432}
\eeq
\beq
\lambda_{3,1,3}=-(q-a_{4,3})=3-q
\label{lam433}
\eeq
\beq
\lambda_{4,3,4}=-(q-a_{4,4})=3+\sqrt{2}-q \ .
\label{lam434}
\eeq
Finally, 
\beq
\lambda_{4,4}=1 \ .
\label{lam44}
\eeq

\subsection{$L_y=4$ M\"obius Strip}

The chromatic polynomial for the M\"obius strip of width $L_y=4$ is \cite{s4}
\beqs
P(4 \times m,Mb) & = & 
c^{(0)} \bigg [ \sum_{j=1}^3 (\lambda_{4,0,j+1})^m 
+ \sum_{j=1}^3 (\lambda_{4,2,j})^m-(\lambda_{4,0,1})^m 
- \sum_{j=1}^5 (\lambda_{4,2,j+3})^m \bigg ] 
\cr\cr
& + & c^{(1)} \bigg [ \sum_{j=1}^5 (\lambda_{4,1,j+4})^m 
-\sum_{j=1}^4 (\lambda_{4,1,j})^m - (\lambda_{4,4})^m \bigg ]
+ c^{(2)}\sum_{j=1}^4 (-1)^j (\lambda_{4,3,j})^m \ . \cr\cr
& & 
\label{ply4mb}
\eeqs
Expressing this in the general form (\ref{pgsummb}), we have
\beqs
P(4 \times m,Mb) & = &
c^{(0)}\bigg [ \sum_{j=1}^{6} (\lambda_{4,0,+,j})^m 
- \sum_{j=1}^{6}(\lambda_{4,0,-,j})^m \bigg ] \cr\cr
& + & c^{(1)} \bigg [ \sum_{j=1}^5 (\lambda_{4,1,+,j})^m 
- \sum_{j=1}^5(\lambda_{4,1,-,j})^m \bigg ] \cr\cr
& + & c^{(2)}\bigg [ \sum_{j=1}^2 (\lambda_{4,2,+,j})^m
- \sum_{j=1}^2 (\lambda_{4,2,-,j})^m \bigg ]
\label{ply4mbgen}
\eeqs
where
\beq
\lambda_{4,0,+,j} = \lambda_{4,0,j+1} \quad {\rm for} \quad 1 \le j \le 3
\label{lam40plusj1to3}
\eeq
\beq
\lambda_{4,0,+,j} = \lambda_{4,2,j-3} \quad {\rm for} \quad 4 \le j \le 6
\label{lam40plusj4to6}
\eeq
\beq
\lambda_{4,0,-,1} = \lambda_{4,0,1}
\label{lam40minusj1}
\eeq
\beq
\lambda_{4,0,-,j} = \lambda_{4,2,j+2} \quad {\rm for} \quad 2 \le j \le 6
\label{lam40minusj2to6}
\eeq
\beq
\lambda_{4,1,+,j} = \lambda_{4,1,j+4} \quad {\rm for} \quad 1 \le j \le 5
\label{lam41plusj1to5}
\eeq
\beq
\lambda_{4,1,-,j} = \lambda_{4,1,j} \quad {\rm for} \quad 1 \le j \le 4
\label{lam41minusj1to4}
\eeq
\beq
\lambda_{4,1,-,5} = \lambda_{4,4}
\label{lam41minus5}
\eeq
\beq
\lambda_{4,2,+,1} = \lambda_{4,3,2} \ , \quad 
\lambda_{4,2,+,2} = \lambda_{4,3,4}
\label{lam42plusj1to2}
\eeq
\beq
\lambda_{4,2,-,1} = \lambda_{4,3,1} \ , \quad
\lambda_{4,2,-,2} = \lambda_{4,3,3} \ .
\label{lam42minusj1to2}
\eeq

\vfill
\eject
\end{document}